\documentclass[10pt,aps,prd,floats,twocolumn,superscriptaddress,nofootinbib,floatfix]{revtex4-2}
\usepackage{amsmath,amssymb, bm,graphicx,booktabs,xcolor}

\usepackage[dvipsnames]{xcolor}
\usepackage{multirow}
\usepackage{verbatim}
\usepackage{indentfirst}
\usepackage[normalem]{ulem}
\usepackage{appendix}
\usepackage{soul}
\usepackage{tikz}
\usepackage{mathtools}
\usepackage{wrapfig}
\usepackage{setspace}
\usepackage{ragged2e}
\usepackage{afterpage}
\usepackage{subcaption}
\usepackage{tablefootnote}
\usepackage{threeparttable}
\usepackage{comment}
\usepackage[hidelinks]{hyperref}
\hypersetup{colorlinks=true, linkcolor=black, citecolor=black, urlcolor=black}

\newcommand{\Hc}{\mathcal H}
\newcommand{\mpl}{m_{\rm pl}}
\newcommand{\kend}{k_{\rm end}}
\newcommand{\Nw}{N_w}
\newcommand{\dd}{\mathrm{d}}
\begin{document}

\title{Gravitational waves and the exit from de Sitter in Inflation without an Inflaton}

\author{Daniele Bertacca}
\thanks{ daniele.bertacca@pd.infn.it}
\affiliation{Dipartimento di Fisica e Astronomia ``Galileo Galilei'', Universit\`a degli Studi di Padova, via Marzolo 8, I-35131 Padova, Italy}
\affiliation{INFN, Sezione di Padova, via Marzolo 8, I-35131 Padova, Italy}
\affiliation{INAF -- Osservatorio Astronomico di Padova, Vicolo dell'Osservatorio 5, I-35122 Padova, Italy}

\author{Raul Jimenez}
\thanks{raul.jimenez@icc.ub.edu}
\affiliation{ICC, University of Barcelona, Mart\'{\i} i Franqu\`es 1, E-08028 Barcelona, Spain}
\affiliation{ICREA, Pg.\ Llu\'{\i}s Companys 23, E-08010 Barcelona, Spain}

\author{Sabino Matarrese}
\thanks{ sabino.matarrese@pd.infn.it}
\affiliation{Dipartimento di Fisica e Astronomia ``Galileo Galilei'', Universit\`a degli Studi di Padova, via Marzolo 8, I-35131 Padova, Italy}
\affiliation{INFN, Sezione di Padova, via Marzolo 8, I-35131 Padova, Italy}
\affiliation{INAF -- Osservatorio Astronomico di Padova, Vicolo dell'Osservatorio 5, I-35122 Padova, Italy}
\affiliation{Gran Sasso Science Institute, Viale F.\ Crispi 7, I-67100 L'Aquila, Italy}

\author{Mariam Abdelaziz}
\thanks{mariamtarekmohamed.abdelaziz-ssm@unina.it}
\affiliation{Scuola Superiore Meridionale, Largo San Marcellino 10, I-80138 Napoli, Italy}
\affiliation{INFN, Sezione di Napoli, Via Cinthia Edificio 6, I-80126 Napoli, Italy}
\affiliation{Department of Astronomy, Space Science and Meteorology, Cairo University, 12613 Giza, Egypt}


\author{Marisol Traforetti}
\thanks{marisol.traforetti@icc.ub.edu}
\affiliation{ICC, University of Barcelona, Mart\'{\i} i Franqu\`es 1, E-08028 Barcelona, Spain}

\date{\today}

\begin{abstract}
In Inflation without an Inflaton (IWI), the primordial curvature perturbation is produced at second order by de Sitter (dS) gravitons. It is the Coulomb-type potential of their gradient stress. Each graviton pair contributes in proportion to the square of the ratio between the internal and external momenta, so the scalar amplitude grows exponentially with the number of e-folds over which pairs contribute. 
We show that this solution, with exact Bunch-Davies time dependence, satisfies all four scalar second-order Einstein equations in pure dS. We then ask whether the tensor modes are enhanced in the same way. Solving the transverse-traceless equation with Bunch-Davies gravitons and the retarded Green function, we find a kernel that stays of order one in the ultraviolet, a factor $(k/p)^2$ below the scalar one, and a hard one-loop correction $\Delta^2_{h,2}/\Delta^2_h\simeq0.04\,\Delta^2_h$. The scalar is fixed by an elliptic constraint, whereas the tensor obeys a wave equation whose source is nonzero even before internal freeze-out. However, the oscillatory Bunch-Davies evolution produces cancellations across the full retarded integral, preventing the $(p/k)^2$ enhancement characteristic of the scalar constraint. 
Next, we follow the enhanced curvature into the radiation era. The curvature that is conserved after dS is that of the hypersurface on which the vacuum energy decays. If the decay is governed by a global clock, carried by the depleting vacuum, that holds each comoving scale on zero-shear time for a fixed number $\Nw$ of e-folds after horizon exit and then lets it run locally, every scale receives the same enhancement. Energy-momentum conservation requires the depletion to feed a non-vacuum sector, which we model as radiation. Solving the second-order equations through a smooth exit and deep into the radiation era, we find $\Delta^2_\zeta=c_w\Delta_h^4e^{\Nw}/20$, with $c_w$ determined by the clock-release prescription, $n_s-1=-4\epsilon$ and $n_T=-2\epsilon$. The spectrum is adiabatic and Gaussian, $\zeta$ is conserved on super-horizon scales to better than $10^{-4}$, and replacing the radiation by another adiabatic fluid changes it by less than $0.2$ per cent. For a direct radiation exit, the data require $31\lesssim\Nw\lesssim50$, or $3\times10^{-6}\lesssim r\lesssim0.036$.
The primordial gravitational-wave background at interferometer frequencies is then $\Omega_{\rm GW}h^2\simeq2\times10^{-21}$--$3\times10^{-17}$, within reach of BBO for $r\gtrsim0.01$ and of an ultimate DECIGO for $r\gtrsim10^{-5}$, while CMB B-mode searches probe $r\gtrsim10^{-3}$.
Beyond $\kend e^{-\Nw}$ the scalar spectrum steepens to $k^{-1}$, which for a sharp release suppresses $\mu$-distortions when $\Nw\gtrsim45$.
\end{abstract}

\maketitle
\section{Introduction}
In standard inflation, a scalar field seeds the curvature perturbation through its quantum fluctuations and ends inflation through its classical roll, while the tensor modes are the vacuum fluctuations of the graviton. 
The \textit{Inflation without an Inflaton} (IWI) framework, recently developed in Refs.~\cite{Bertacca:2024zfb,Traforetti:2025cax,Abdelaziz:2026qyk,Jimenez:2026rqn}, removes the scalar field. De Sitter (dS) space is described as a coherent state of gravitons~\cite{Dvali:2013eja,Dvali:2017eba}, the vacuum energy ends through the quantum break of that state, and the curvature perturbation is produced by the gravitons themselves at second order. 

Hence, there is no first-order scalar source. For tensor perturbations in a dS background, we consider the Bunch-Davies state of the graviton, with $\langle\chi_{ij}\rangle=0$ and $\langle\chi_{ij}\chi_{kl}\rangle\neq0$, and the curvature perturbation is the second-order response of the metric to it, $\delta^{(1)}G^\mu{}_\nu[g^{(2)}]=-\delta^{(2)}G^\mu{}_\nu[\chi,\chi]$, where $g^{(2)}$ is the second-order metric. At this order, the Einstein equations fix the scalar potential through a constraint, and the potential is the Coulomb-type potential of the gravitons' gradient stress, $\phi\sim\nabla^{-2}(\partial\chi\,\partial\chi)$. A pair of gravitons with momenta $\mathbf p_1,\mathbf p_2$ and total momentum $\mathbf k$ contributes an amount that grows as $(p/k)^2$, once both modes have frozen. Summing over pairs up to a maximum ratio $x_{\max}=e^{N_{\rm obs}}$ gives $A_s\propto\Delta_h^4e^{N_{\rm obs}}$~\cite{Traforetti:2025cax}, where $N_{\rm obs}$ is the number of observed e-folds and $\Delta_h^2$ is the tensor power spectrum. Ref.~\cite{Traforetti:2025cax} takes the same ratio $x_{\rm max}$ for every scale, and interprets the fixed ratio as a nonlinear coupling between modes, either between phases of the graviton condensate or through interactions between particles and gravitons. A slow drift of the Hubble rate, which can be tied to the depletion of the coherent state~\cite{Dvali:2017eba,Gomez:2021jcl,Jimenez:2026rqn,bridge}, gives the small red tilt $n_s-1=-4\epsilon$~\cite{Jimenez:2026rqn}.

If a pair of gravitons can produce an enhanced scalar, the same second-order equations might also produce an enhanced second-order tensor, with a tensor-to-scalar ratio $r$ of order unity that would rule the scenario out. We find that this does not happen. The TT sector obeys a wave equation rather than an elliptic scalar constraint, and the oscillatory Bunch-Davies evolution suppresses the retarded tensor response without requiring the source to vanish before internal freeze-out, leaving a small second-order correction that is finite in the ultraviolet.

We then study what happens to the scalar when dS ends. The second-order response is computed in pure GR$+\Lambda$, whereas the Hubble rate drifts and the vacuum energy eventually decays. Such a vacuum energy cannot be conserved by itself, and the Bianchi identity forces it to feed a non-vacuum sector. In pure dS there is no matter whose density or velocity could define the usual conserved curvature $\zeta$, and what IWI computes is the curvature of the zero-shear slices. After the exit, the conserved $\zeta$ of the radiation era is the curvature of the hypersurface on which the vacuum energy decays~\cite{Israel:1966rt,Deruelle:1995kd}, so the clock that ends de Sitter decides what survives (Sec.~\ref{sec: Clock, sink and the physical curvature}). In standard inflation, this clock is the inflaton, and $\zeta$ is the number of e-folds each patch accumulates before the field reaches its final value~\cite{Starobinsky:1985ibc,Sasaki:1995aw,Lyth:2004gb}.

We show that in IWI the enhanced scalar survives, together with the observed tilt, if the vacuum decays on a global clock that keeps each comoving scale on zero-shear time for a fixed number of e-folds after the scale leaves the horizon. After that, the clocks keep their readings and tick with local proper time. Since the window is the same for every scale, the spectrum is scale invariant, and the drift of the Hubble rate across the window supplies the tilt. Such a clock is necessarily non-local; it is carried by the depleting vacuum energy, and we model the sector that receives the energy as a passive sink. For adiabatic sinks, we find that a change in the equation of state moves the result by less than $0.2$ per cent, and the local entropy and anisotropic responses we have tested rescale the amplitude, but do not produce the window. Any microscopic completion has to produce the window through the response of the vacuum and the sector it feeds taken together.

We also clarify the relation with Refs.~\cite{Bertacca:2024zfb,Traforetti:2025cax}. The solution $\psi_2=0$, $\phi_2=\mathcal F_\chi/4$ of Ref.~\cite{Bertacca:2024zfb} is the late-time limit of the exact Bunch-Davies solution, which satisfies all four scalar Einstein equations in pure GR$+\Lambda$ (Appendix~\ref{app:consistency}). The second-order equations of Ref.~\cite{Bertacca:2024zfb} in pure GR$+\Lambda$ are consistent without additional matter. An apparent overconstraint, which would make a fluid term necessary for consistency~\cite{Traforetti:2025cax}, arises only when time derivatives are set to zero before the late-time limit is taken. The fluid of Ref.~\cite{Bertacca:2024zfb}, which is the averaged stress of the sub-horizon gravitons, is therefore not needed for consistency. In the present construction, the drift and the exit come instead from the depleting vacuum and the sector it feeds, and the physical curvature is that of the surface on which the vacuum decays. Our Einstein tensor reproduces the source kernels of Ref.~\cite{Traforetti:2025cax} and, to 2 per cent, its amplitude (Sec.~\ref{sec: The graviton-induced curvature}). The window used here is a dynamical realisation of the fixed-ratio prescription of Ref.~\cite{Traforetti:2025cax}, with $\Nw$ in place of $N_{\rm obs}$, and it has to respect the end of dS (Sec.~\ref{sec: Clock, sink and the physical curvature}).

In section~\ref{sec: The graviton-induced curvature} we present the setup of the second-order calculation. Section~\ref{sec: Tensor modes are not enhanced} studies the second-order tensors, while section~\ref{sec: The exit from de Sitter} describes the exit, the role of the depletion sector and the evolution of the curvature through the transition. In section~\ref{sec: Observational consequences} we discuss the observational consequences, and we conclude in section~\ref{sec: Discussion}. The technical material is collected in the Appendixes.

\section{The graviton-induced curvature}
\label{sec: The graviton-induced curvature}
\subsection{Second-order metric and normalisation}

We write the metric to second order as
\begin{align}
d s^2=a^2(\eta)\Big\{&-(1+\psi)\dd\eta^2+2\partial_iB\,\dd x^i\dd\eta\nonumber\\
&+\big[(1-\phi)\delta_{ij}+2\partial_i\partial_jE\nonumber\\
&+\chi_{ij}+h^{(2)}_{ij}\big]\dd x^i\dd x^j\Big\}.
\label{eq:metric}\end{align}
Here, $\chi_{ij}$ is the first-order TT graviton, $h^{(2)}_{ij}$ is the second-order TT tensor, and $\phi,\psi,B,E$ are second-order scalars. There are no first-order scalars. In the Poisson gauge, $B=E=0$, the IWI potential is $\phi_{\rm IWI}=\phi/2$.

In de Sitter (dS), the scale factor reads $a(\eta)=-1/(H\eta)$, where $H$ is the energy scale of inflation. The gravitons are the Bunch-Davies vacuum fluctuations, and every second-order field below is a response to them. More precisely, the tensor modes are $\chi_{ij}\propto e_{ij}\,\mathcal T(p\eta)$ with $\mathcal T(x)=(1+ix)e^{-ix}$, normalised so that $\mathcal T\to1$ outside the horizon, and the tensor power per logarithmic interval is $\Delta_h^2=(16/\pi)(H/\mpl)^2$, with $\mpl$ the Planck mass. We use the canonical normalisation $\langle\chi_{ij}\chi_{ij}\rangle'=P_h$, with polarisation tensors satisfying $e:e^*=2$ and a power $P_h/4$ per helicity. Refs.~\cite{Bertacca:2024zfb,Traforetti:2025cax,Abdelaziz:2026qyk} use $P_h/2$ per helicity, which multiplies every scalar power spectrum by a factor of 4; hence, at fixed $r$ their window $\Nw$, introduced below, is shorter by $\ln4$.

We compute the mixed Einstein tensor $G^\mu{}_\nu$ exactly to second order, with a first-order field made of two plane waves of generic momenta and polarisations,
\begin{equation}
\chi_{ij}=E^{(1)}_{ij}f_1(\eta)e^{i\mathbf p_1\cdot\mathbf x}+E^{(2)}_{ij}f_2(\eta)e^{i\mathbf p_2\cdot\mathbf x}.
\end{equation}
where $\mathbf p_a$ are the internal momenta, $E^{(a)}_{ij}=e^{\lambda_a}_{ij}(\hat{\mathbf p}_a)$ are the
TT tensors encoding the polarisation and amplitude of each helicity $\lambda_a$, while $f_a(\eta)\equiv\mathcal T(p_a\eta)=(1+ip_a\eta)e^{-ip_a\eta}$ are the Bunch-Davies mode functions, with $p_a=|\mathbf p_a|$ and $a=1,2$.

The second-order fields sit at $\mathbf k=\mathbf p_1+\mathbf p_2$, and we read their equations off the bilinear cross term. We denote as $Q^\mu{}_\nu$ the graviton-bilinear part of $a^2G^\mu{}_\nu$. The result reproduces the second-order source kernels of Ref.~\cite{Traforetti:2025cax} to $2\times10^{-14}$, so the two describe the same second-order system. Moreover, its tensor sector reproduces Maldacena's graviton bispectrum exactly, in shape, polarisation structure and amplitude~\cite{Maldacena:2002vr}; its TT source agrees with that of Ref.~\cite{Chen:2025fcd} to $6\times10^{-15}$; and under $x^\mu\to x^\mu+\xi^\mu$ the full four-scalar operator changes by exactly $-\mathcal L_\xi\bar G^\mu{}_\nu$. The second-order equations are listed in Appendix~\ref{app: Second-order equations}.

\subsection{The zero-shear potential and its enhancement}
\label{sec: The zero-shear potential and its enhancement}
In vacuum, the scalar sector is algebraic at every instant. Combining the $00$ and $0i$ equations gives a comoving Poisson equation, and the traceless part of the $ij$ equation gives the anisotropic-stress relation,
\begin{align}
k^2\phi&=-Q^0{}_0+3i\Hc\,\hat{\mathbf k}\cdot\mathbf Q^0/k,\label{eq:poisson}\\
\phi-\psi&=T\equiv\frac{3k_ik^jQ^i{}_j}{k^4}-\frac{Q^l{}_l}{k^2}\quad(B=E=0).\label{eq:traceless}
\end{align}

These constraints determine the inhomogeneous scalar response at $k\neq0$. The exactly homogeneous mode is governed instead by the spatially averaged Einstein equations, as discussed in Appendix~\ref{app: The homogeneous scalar mode and graviton backreaction}.

At late times $\psi\to0$ and $\phi$ tends to $\mathcal F_\chi/4$, where $\mathcal F_\chi$ is the inverse-Laplacian functional of the frozen gravitons given in Ref.~\cite{Bertacca:2024zfb}. The late-time limit is thus the solution $\psi_2=0$, $\phi_2=\mathcal F_\chi/4$ identified there. The exact solution satisfies all four scalar Einstein equations (see Appendix~\ref{app:consistency}). Once the internal modes have frozen, we get $Q\sim p^2\chi\chi$ and the inverse Laplacians give a kernel $\propto(p/k)^2$, so the curvature of the zero-shear slices,
\begin{equation}
\zeta_{\rm P}=-\tfrac12\phi-\tfrac1{12}\chi_{ij}\chi_{ij},
\label{eq:zetaP}\end{equation}
grows as $(p/k)^2$ per pair \footnote{The second-order curvature variable of Ref.~\cite{Bertacca:2024zfb} and the fully geometric non-linear curvature \eqref{eq:zetaP} used here agree at quadratic order once normalisations are matched. Beyond $-\phi/2$, the local-volume variable of Ref.~\cite{Bertacca:2024zfb} carries the tensor term $-\chi_{1ij}\chi_1^{ij}/6$, which in the field normalisation of the present work is the $-\chi_{ij}\chi_{ij}/12$ of Eq.~\eqref{eq:zetaP}. The uniform-density contribution $-\Hc\,\delta^2\rho/\bar\rho'$ of Ref.~\cite{Bertacca:2024zfb}, which vanishes in exact dS, is what defines the physical curvature once the vacuum drifts and decays: Section~\ref{sec: Clock, sink and the physical curvature} and Appendix~\ref{app: Uniform density versus zero shear} show it is negligible after a zero-shear exit, while with the window clock it carries the clock offset $\delta N_\Sigma$ of Eq.~\eqref{eq:junction}.}. 
In pure dS, $\zeta_{\rm P}$ is a well-defined geometric quantity, but it is not by itself the observable, since there is no matter whose density could define uniform-density slices (see the discussion in section~\ref{sec: Clock, sink and the physical curvature}).

Writing a second-order field $X$ as
\begin{eqnarray}
\label{eq: second order field}
X(\mathbf k)=\sum_{\lambda_1\lambda_2}\int \dd\mathbf p_1 \;K_X\,h_{\lambda_1}(\mathbf p_1)h_{\lambda_2}(\mathbf{k}-\mathbf p_1),
\end{eqnarray}
with $h_\lambda$ the helicity amplitudes and $K_X$ the symmetrised pair kernel, its power spectrum is
\begin{eqnarray}
\Delta^2_X(k)=\frac{\Delta_h^4}{64}\int_0^{\infty}\dd x\;g_X(x),\nonumber\\ g_X=\int_{|1-x|}^{1+x}\dd y\,\frac{\sum_{\lambda_1\lambda_2}|K_X|^2}{x^2y^2},
\label{eq:loop}
\end{eqnarray}
where $x=p_1/k$, $y=p_2/k=|\mathbf{k}-\mathbf p_1|/k$ are dimensionless variables, and $\dd^3p=2\pi k^3xy\,\dd x\,\dd y$. The normalisation of Eq.~\eqref{eq:loop} follows from Wick's theorem with two contractions. For the zero-shear curvature $\zeta$, we obtain $g_\zeta\to16/5$ at large $x$ (see Fig.~\ref{fig:kernels}). Every pair contributes equally, and
\begin{equation}
\Delta^2_\zeta=\tfrac1{20}\Delta_h^4x_{\rm max},
\label{eq:As}\end{equation}
where $x_{\rm max}$ is the largest $x$ that contributes \footnote{In the normalisation of Refs.~\cite{Bertacca:2024zfb,Traforetti:2025cax,Abdelaziz:2026qyk}, with $P_h/2$ per helicity, this reads $A_s=\Delta_h^4e^{N_{\rm obs}}/5$ for $x_{\rm max}=e^{N_{\rm obs}}$, and it reproduces the $H(N_{\rm obs})$ and $r(N_{\rm obs})$ of Ref.~\cite{Traforetti:2025cax}. The kernels reported in Appendix~C of Ref.~\cite{Traforetti:2025cax} integrate to $3.141$ instead of $16/5$, the asymptotic difference coming from the $\mathcal K^{(AB)}$ term. This is a $1.8$ per cent difference in amplitude, which moves $H$ by $0.5$ per cent at fixed $A_s$ and $\Nw$ by $0.02$ e-folds at fixed $r$, far below the other uncertainties of the problem. The two calculations therefore agree on the amplitude as well as on the kernels.}. 
The ultraviolet dominates because at fixed time the graviton stress is a local bilinear dominated by the shortest frozen modes, so on large scales it behaves as white noise and $\zeta_{\rm P}$ is its Poisson potential. How large $x_{\rm max}$ is, and whether it is the same for every $k$, is decided by the exit from dS, as we will discuss in section~\ref{sec: The exit from de Sitter}.

\section{Tensor modes are not enhanced}
\label{sec: Tensor modes are not enhanced}
\subsection{The second-order tensor equation}
\label{sec: The second-order tensor equation}
The TT projection of the second-order $ij$ equation gives
\begin{align}
&h^{(2)\prime\prime}_{ij}+2\Hc h^{(2)\prime}_{ij}+k^2h^{(2)}_{ij}=S_{ij},\nonumber\\
&S_{ij}\equiv-2\Lambda_{ij,lm}Q^l{}_m=S^{FF}_{ij}f_1f_2+S^{dd}_{ij}f_1'f_2',
\label{eq:TT}
\end{align}
with $\Lambda_{ij,lm}$ the TT projector, which removes exactly the trace and longitudinal parts of $Q_{ij}$ that carry the $\nabla^{-2}$ of Eq.~\eqref{eq:traceless}. The tensors $S^{FF}$ and $S^{dd}$ are constant, built from momenta and polarisations alone, and the source does not contain $\Hc$-dependent terms, with all the time dependence of the source in the two mode-function products. 
The $f_1'f_2'$ term, $S^{dd}=\Lambda(E^{(1)}E^{(2)}+E^{(2)}E^{(1)})$, is a pure field redefinition, since in exponential variables, $g_{ij}=a^2(e^\gamma)_{ij}$, the source is purely gradient, as in Maldacena's cubic action~\cite{Maldacena:2002vr}.

\subsection{The causal solution}
\label{sec: The causal solution}
The retarded Green function of the operator in Eq.~\eqref{eq:TT} is
\begin{equation}
G_k(\eta,\eta')=\frac{\mathcal T(k\eta')\mathcal T^*(k\eta)-\mathcal T(k\eta)\mathcal T^*(k\eta')}{2i\,k^3\eta'^2}.
\label{eq:green}
\end{equation}
For Bunch-Davies gravitons, $f_a(\eta)\equiv\mathcal T(p_a\eta)$, and the solution of Eq.~\eqref{eq:TT} that vanishes in the far past, at late time, is
\begin{align}
h^{(2)}(\mathbf k,0)&=S^{FF}J_{FF}+S^{dd}J_{dd},\nonumber\\
J_{FF}&=-\frac{P^2+2p_1p_2-k^2}{(P^2-k^2)^2},\quad
J_{dd}=\frac{2p_1^2p_2^2}{(P^2-k^2)^2},
\label{eq:J}
\end{align}
with $P=p_1+p_2$ (Appendix~\ref{app: Tensor time integrals}). A direct numerical integration along a rotated contour reproduces these closed forms to $10^{-10}$. For $p_{1,2}\gg k$, they tend to $J_{FF}\to-3/(8p^2)$ and $J_{dd}\to1/8$, so the response kernel is of order one, a factor $(k/p)^2$ below the zero-shear scalar.

This absence of the enhancement comes from oscillatory cancellations across the full retarded Bunch-Davies integral. Although the TT source is nonzero before internal freeze-out, the combined contributions from the subhorizon and frozen regimes yield an $O(1)$ tensor response, so $h^{(2)}\sim\chi\chi$.
The fact that the scalar acquires the $(p/k)^2$ enhancement follows because Eqs.~\eqref{eq:poisson} and \eqref{eq:traceless} are constraints that hold at every instant, whereas Eq.~\eqref{eq:TT} is a wave equation whose response is built up over time. 
To see this more clearly, suppose instead that the source were treated as frozen from the moment the external mode crosses the horizon, i.e. $|\eta|=1/k$, even though the internal modes are still oscillating at that time. The solution would then be $h^{(2)}\simeq(1-\sin1)S^{FF}/k^2$, with constant mode $S^{FF}/k^2$; both carry the same $(p/k)^2$ as the scalar that is removed by the oscillatory evolution. 

\begin{figure}[t]\centering
\includegraphics[width=\columnwidth]{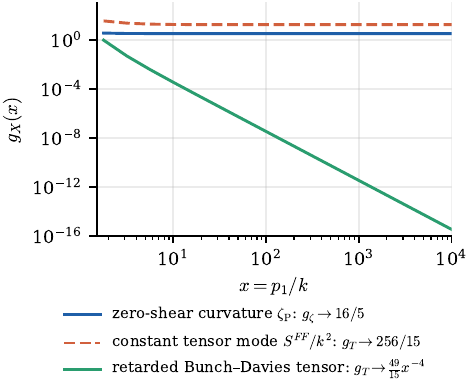}
\caption{We plot $g_X(x)$, which is the helicity-summed, $y$-integrated squared kernel of Eq.~\eqref{eq:loop}, as a function $x=p_1/k$. For the zero-shear curvature $\zeta_{\rm P}$, $g_\zeta$ tends to a constant (solid blue line), so its power grows with the number of contributing graviton pairs. The constant tensor mode $S^{FF}/k^2$ behaves the same way, so $g_T$ tends to a constant value (dashed orange line). On the other hand, the retarded Bunch-Davies tensor decays as $g_T\to\tfrac{49}{15}x^{-4}$ (solid green line), and its loop converges.}
\label{fig:kernels}
\end{figure}

\subsection{Power spectrum}
\label{sec:tensorpower}
Inserting Eq.~\eqref{eq:J} into Eq.~\eqref{eq:loop} gives $g_T\to\tfrac{49}{15}x^{-4}$ (see Fig.~\ref{fig:kernels}), so the loop converges and depends neither on the ultraviolet cutoff nor on the duration of dS. Over the hard region $p_1,p_2>k$, we find $\int g_T\,\dd x=2.33$, dominated by $x\lesssim4$, and giving a second-order tensor power
\begin{equation}
\frac{\Delta^2_{h,2}}{\Delta^2_h}\bigg|_{\rm hard}=\frac{2.33}{64}\,\Delta^2_h=0.036\,\Delta^2_h\lesssim3\times10^{-12}.
\label{eq:tensorloop}
\end{equation}
Considering the constant mode $S^{FF}/k^2$ would instead give $g_T\to256/15$ (see Fig.~\ref{fig:kernels}) and $\Delta^2_{h,2}=\tfrac{16}{3}A_s$, i.e. $r\simeq5$, and the solution with the source switched on at $k|\eta|=1$ is smaller by $(1-\sin1)^2\simeq0.025$ and gives $r\simeq0.13$, still above the bound $r<0.036$~\cite{BICEP:2021xfz}. Only the full retarded Bunch-Davies evolution brings the tensor down to Eq.~\eqref{eq:tensorloop}.

The soft and collinear parts of $\langle h^{(2)}h^{(2)}\rangle$ contain the usual phase-accumulation poles at $P=k$, which are not specific to IWI. We expect them to cancel against $\langle\chi h^{(3)}\rangle$ in the complete one-loop result, as the same poles cancel in the tree-level bispectrum once the operator ordering is consistent. We have not computed $\langle\chi h^{(3)}\rangle$; power counting bounds the remainder by $\mathcal O(\Delta_h^2\ln(\kend/k))$.

With no first-order scalars, $h^{(2)}_{ij}$ is invariant under second-order gauge transformations, since second-order generators only shift $\phi,\psi,B,E$ and the vectors. The tensors therefore do not depend on the slicing on which dS ends, and super-horizon tensors cross the transition unchanged. 
The tree-level tensor bispectrum is exactly that of GR, with $f^{+++}_{\rm eq}\simeq1.69$ and $f^{++-}_{\rm eq}\simeq0.02$ in the helicity basis, which corresponds to a fractional non-Gaussianity $f\Delta_h/2\lesssim10^{-5}$. Once the enhanced scalar exists, it sources tensors in the radiation era only at order $A_s^2$~\cite{Matarrese:1997ay,Mollerach:2003nq,Ananda:2006af,Baumann:2007zm,Kohri:2018awv}, and tensor-induced tensors in the radiation era are of order $\Delta_h^4$~\cite{Chen:2025fcd}. Both are negligible as we will discuss in Sec.~\ref{sec: The gravitational-wave background today}.

\subsection{Momentum window versus temporal support}
\label{sec: Momentum window versus temporal support}
The scalar mechanism of Sec.~\ref{sec: The exit from de Sitter} restricts the pairs that count to $p\lesssim k\,e^{\Nw}$. Imposing the same momentum window on the tensor loop changes nothing: with the retarded Bunch-Davies kernel the part of the loop beyond $x_{\rm max}=e^{\Nw}$ is a fraction $0.47\,x_{\rm max}^{-3}$ of Eq.~\eqref{eq:tensorloop}, below $10^{-34}$ for any $\Nw\gtrsim26$. The momentum window is therefore irrelevant to the tensors.

In fact, the tensor result depends only on the time over which the TT source acts. The tensor is small because the internal gravitons oscillate and average out, while the external mode is already outside the horizon. A source acting from the external crossing would instead give a constant $g_T$ and $r$ in the range $0.1$--$5$ (Sec.~\ref{sec:tensorpower}). Our tensor result therefore holds provided the depletion acts as a scalar clock and adds no long-range, correlated spin-2 response that anticipates or replaces the Bunch-Davies evolution of the TT source.

In the model of Sec.~\ref{sec: The exit from de Sitter} this condition is met automatically. The clock perturbation is isotropic, $\delta T_v{}^\mu{}_\nu=-\delta\rho_v\delta^\mu{}_\nu$, and the radiation sink is a perfect fluid whose TT stress starts at fourth order. A free-streaming sink would also carry a first-order TT stress that damps the tensor legs, as neutrinos do in the radiation era~\cite{Weinberg:2003ur}, but with an energy fraction below $10^{-2}$ the effect is below a per cent. 
A scalar anisotropic stress of the sink, on the other hand, has no TT projection and enters only the scalar sector. The scalar perturbations that the clock leaves behind source tensors only at second order in themselves, at the scalar-induced level $A_s^2$ (see Sec.~\ref{sec: The gravitational-wave background today}), and cross terms between the enhanced scalar and the gravitons correct the tensor power at relative order $A_s$. 
A microscopic depletion carrying long-range spin-2 correlations would be a new source of gravitational waves, and the tensor calculation would have to be redone with it included. Within the present model, no such source is present, and the gravitational waves of IWI are the first-order vacuum gravitons, with a tensor tilt of $n_T=-2\epsilon$, where $\epsilon$ parametrizes the slow drift in the Hubble rate.

\section{The exit from de Sitter}
\label{sec: The exit from de Sitter}
\subsection{What the exit clock has to do}
\label{sec: What the exit clock has to do}
The vacuum energy converts into radiation on a hypersurface $\Sigma$ defined by some clock. For super-horizon modes, the junction conditions for a transition from $w=-1$ to $w=1/3$~\cite{Israel:1966rt,Deruelle:1995kd} give
\begin{equation}
\zeta_+=\zeta_{\rm P}+\delta N_\Sigma,
\label{eq:junction}\end{equation}
where $\delta N_\Sigma=\Hc\,\delta\eta_\Sigma$ is the delay of $\Sigma$ relative to the zero-shear slice. The value $\zeta_+$ is the curvature of $\Sigma$ itself, and it is conserved afterwards because the radiation era starts as a single adiabatic fluid. A smooth transition computed without junction conditions gives the same result (see Sec.~\ref{sec: Solving through the transition}).

We split the exit time of each patch into its proper time and an offset, $\zeta_+=\zeta_\tau+H\Delta t$. Considering the proper-time part, we find that it is local and bounded. Along the comoving geodesics of dS, the vacuum Einstein equations give exactly $R_{\mu\nu}u^\mu u^\nu=-\Lambda$, so Raychaudhuri's equation reads $\dot\theta=-\theta^2/3-\sigma_{ij}\sigma^{ij}+\Lambda$. Writing $\theta=3H+\delta\theta$, the local number of e-folds at fixed proper time is $\delta N_V=-(6H)^{-1}\int\sigma^2\dd t$. For a graviton pair $\sigma^2=\tfrac14\dot\chi_{ij}\dot\chi_{ij}$ and with $f'(\eta)=p^2\eta\,e^{-ip\eta}$, one obtains
\begin{equation}
\delta N_V=-\frac12\,E^{(1)}:E^{(2)}\,\frac{p_1^2p_2^2}{(p_1+p_2)^4},
\label{eq:raychaudhuri}\end{equation}
which tends to a constant for $p\gg k$. The curvature $\zeta_\tau$ of the constant-proper-time slice differs from this volume term by an anisotropic piece, $-k^2E/3$ in synchronous gauge, which is also local and bounded. For $(x,y)=(10,10.4)$ in exact dS $\delta N_V=-0.062$ and $\zeta_\tau=-0.138$.

The whole enhancement is thus in the offset $H\Delta t$, and a clock that carries it has to follow zero-shear time. Such a clock is non-local, because any clock built from local physics ends dS on a surface whose curvature is the bounded local term. In standard inflation, the inflaton plays this role, since its fluctuation $\delta\phi$, generated at horizon exit and carried locally, makes different patches reach the end at different times.

Finally, the clock must also give a scale-invariant spectrum. By Eq.~\eqref{eq:As}, the spectrum is set by how many pairs each scale collects, and a scale collects pairs for as long as its patches stay synchronised on zero-shear time. For the amplitude to be the same for every scale, every scale must collect pairs for the same number of e-folds, counted from its own horizon exit.

\subsection{A zero-shear clock with a finite window}
\label{sec: A zero-shear clock with a finite window}
\begin{figure}[t]\centering
\includegraphics[width=\columnwidth]{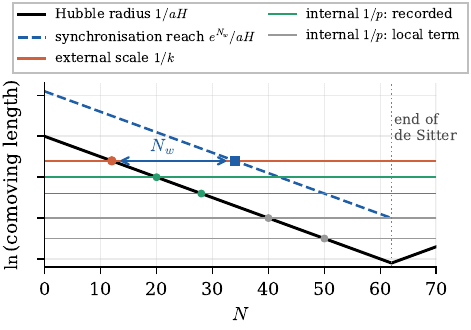}
\caption{We show the picture of the mechanism by plotting the comoving scale as a function of the number of e-folds $N$. A comoving scale $1/k$ (solid orange line) leaves the Hubble radius (solid black line) and remains synchronised on zero-shear time within the synchronisation reach $e^{\Nw}/aH$ (dashed blue line), where $\Nw$ is the window's number of e-folds. The moment of horizon crossing is shown with circles, while the release happens $\Nw$ e-folds later and it is shown with a square. Internal modes $1/p$ that freeze before this release are recorded in the clock offset (solid green line). Internal modes that freeze later contribute only a bounded local term (solid gray lines). The dotted gray line shows the end of the de Sitter phase.}
\label{fig:schematic}\end{figure}

We have seen in Sec. \ref{sec: What the exit clock has to do} the requirements on the clock, i.e. a non-local, zero-shear clock with a per-scale window.
Here, we take a global clock that holds each comoving scale on zero-shear time while the scale is less than $\Nw$ e-folds outside the horizon, $\ln(\Hc/k)<\Nw$. After that, the clocks keep their readings and tick with local proper time, and dS ends when they reach a common value. Equivalently, the synchronising influence reaches a fixed physical distance of $e^{\Nw}$ Hubble radii at every moment (see Fig.~\ref{fig:schematic}), which corresponds to $10^{13}$--$10^{22}$ Hubble radii across the allowed range. We postulate this clock and identify it with the break of the dS coherent state as a whole~\cite{Dvali:2017eba,Gomez:2021jcl}. The depletion fixes the common reading at which dS ends, and the coherence of the state across super-horizon distances fixes the window. This is a property of the state as a whole and is distinct from the coherence of individual graviton pairs, which become classical within $9$--$23$ e-folds of horizon exit~\cite{bridge}.

Let $\xi$ be the time shift from the zero-shear slices to the clock slices, $\eta\to\eta+\xi$, so that the clock slices have curvature $\zeta_{\rm P}-\Hc\xi$ and $\delta N_\Sigma=-\Hc\xi$ in Eq.~\eqref{eq:junction}. Then, the clock is defined by 
\begin{equation}
\text{synchronised: }\xi=0;\qquad\text{free-running: }\xi'=\tfrac12\psi-\Hc\,\xi\,.
\label{eq:clock}
\end{equation}
The free-running condition is that the clock reads proper time along the zero-shear normals, $\tilde\psi=\psi-2(\xi'+\Hc\xi)=0$. We spread the release smoothly over $\Delta N_{\rm r}$ e-folds around $\ln(\Hc/k)=\Nw$. Since the vacuum energy is uniform on the clock slices, in Poisson gauge it carries $\delta\rho_v=\bar\rho_v'\,\xi$ and $\delta p_v=-\delta\rho_v$.

Considering a single pair, it builds its zero-shear offset while its internal modes freeze, at $\Hc\simeq p$. If the scale $k$ is still synchronised at that time, then the clock carries the whole offset forward. On the other hand, if it is not synchronised, the free-running clock integrates the lapse, which cancels the offset and leaves only the local term. Only pairs with $p\lesssim k\,e^{\Nw}$ therefore contribute. By the dilation symmetry of dS, the kernel depends on $x$, $y$ and $\Nw$ but not on $k$, so every scale receives the same enhancement. This requires each clock to keep, after the release, the offset it has acquired instead of resetting to local time.

\subsection{Clock, sink and the physical curvature}
\label{sec: Clock, sink and the physical curvature}
The calculation of section~\ref{sec: The graviton-induced curvature} is pure GR$+\Lambda$, whereas in the cosmology we want to describe the Hubble rate drifts, the vacuum depletes, and dS ends. A vacuum component $T_v{}^\mu{}_\nu=-\rho_v\delta^\mu{}_\nu$ satisfies $\nabla_\mu T_v{}^\mu{}_\nu=-\partial_\nu\rho_v$ identically, so if $\rho_v$ changes the Bianchi identity requires a second sector with
\begin{equation}
\nabla_\mu T_{\rm eff}{}^\mu{}_\nu=\partial_\nu\rho_v\equiv J_\nu \,.
\label{eq:transfer}
\end{equation}
Within GR, the minimal content of the model is therefore a depleting vacuum together with a non-vacuum sector that absorbs its energy. Because $p_v=-\rho_v$, the transfer is not a free function, but it is fixed by the slices on which $\rho_v$ is uniform, with no momentum transfer in their frame. A different $J_\nu$ would require the depleting component to depart from $p_v=-\rho_v$ at the perturbative level, for instance by carrying momentum or anisotropic stress. 
The sink is a product of the depletion, analogous to the reheating products of standard inflation but produced continuously at order $\epsilon$. It is the non-vacuum enthalpy that any drift of $H$ requires, since $-2\dot H=\rho+p$, and in the background it plays the role that $\dot\phi^2$ plays in standard inflation, but, unlike the inflaton, it is not the source of the curvature perturbation. We model the sink as a passive fluid \footnote{This closure is an assumption, not a consequence of the Bianchi identity, and a microscopic treatment may give the produced sector a density, entropy, stress, and memory response of its own.}. 
The homogeneous quadratic stress of the Bunch-Davies gravitons is distinct from this depletion-generated sector. Its finite-window contribution is of order $\Delta_h^2$, far below the background $\epsilon$ required by the observed tilt. The corresponding homogeneous backreaction and its renormalized interpretation are examined in Appendix~\ref{app: The homogeneous scalar mode and graviton backreaction}.

The clock and the sink play different roles. A slicing can be built from any background scalar $X$ with $\bar X'\neq0$, and its curvature is
\begin{equation}
\zeta_X=\zeta_{\rm P}-\Hc\,\frac{\delta X}{\bar X'} \,.
\label{eq:zetaX}
\end{equation}
For a perfect fluid with no exchange $\bar\rho'=-3\Hc(\rho+p)$, and this reduces to the familiar $\zeta_{\rm P}+\delta\rho/[3(\rho+p)]$.
Pure $\Lambda$ has $\bar\rho'=0$ and defines no clock, which is why $\zeta_{\rm P}$ alone is not observable. The depleting vacuum, on the other hand, has $\bar\rho_v'\neq0$, so its uniform slices define one even though $\rho_v+p_v=0$. In our construction, these are the clock slices of Eq.~\eqref{eq:clock}, with $\delta\rho_v=\bar\rho_v'\xi$. This identifies the slices on which the clock is read, but it does not give the law by which they are synchronised, released, and keep their memory, which we describe in Sec.~\ref{sec: A zero-shear clock with a finite window}, and the post-exit curvature depends on that law. 
The sink cannot take over the clock's role. After freeze-out, its uniform-density slices converge to the zero-shear ones, as do those of the total density (Appendix~\ref{app: Uniform density versus zero shear}), with an offset larger by a factor $2/\epsilon$, although both decay as $p/aH$. Reading the curvature on the sink's slices would therefore record every pair up to $\kend$ and give $\Delta^2_\zeta\propto\kend/k$.

After the exit the matter is an ordinary fluid, and the curvature on uniform total density,
\begin{equation}
\zeta=-\tfrac12\phi-\tfrac1{12}\chi_{ij}\chi_{ij}+\frac{\delta\rho}{3(\rho+p)},
\label{eq:zetaUD}
\end{equation}
is defined and conserved. This is the quantity to which $A_s$ refers. It equals the curvature of the clock surface, hence $\zeta_{\rm P}$ for a zero-shear exit, and it follows from the clock dynamics, the transfer, and the matching (see Appendix~\ref{app: Uniform density versus zero shear}). 
During quasi-dS, while the clock is synchronised and once the internal modes have frozen, the density term $\delta\rho/[3(\rho+p)]$ in Eq. \eqref{eq:zetaUD} reduces to the velocity term $\Hc v=-\psi/6$ and decays as $p/aH$. After a zero-shear exit, the uniform-density curvature equals $\zeta_{\rm P}$ to about $10^{-3}$, and with the window clock the post-exit value does not depend on when the exit happens, to $10^{-6}$.

We consider radiation as a natural sink. With $q\equiv-\dot\rho_v>0$,
\begin{equation}
\dot\rho_r+4H\rho_r=q,\qquad \epsilon\equiv-\frac{\dot H}{H^2}=\frac32\,\frac{\rho+p}{\rho}=\frac{2\rho_r}{\rho},
\label{eq:sinkbg}
\end{equation}
so a steady drift has $\rho_r=\epsilon\rho/2$. 
For a sink with energy fraction $\Omega_e$ and equation of state $w_e$ the same relation reads $\epsilon=\tfrac32\Omega_e(1+w_e)$. Acceleration ends at $\epsilon=1$, so $\rho_r=\rho/2$, which requires the transfer to grow to $q\sim H\rho$. A steady depletion does not get there by itself. In fact, for $q=gH\rho_v$ with constant $g<2$ the background approaches a fixed point with $\epsilon=g/2$ for any sink with $w_e>-1/3$, and inflation continues. 
The exit therefore needs a change in the rate or in the state, which we impose on the clock slices over $\Delta N=0.3$. 
If the radiation is replaced by a fluid with constant $w$ and rest-frame sound speed $c_s^2$, the velocity term becomes $\Hc v=-\psi/(4+6c_s^2)$, independent of $w$. 
For adiabatic sinks ($c_s^2=w$) and for $c_s^2>w$ the post-exit $|\zeta|$ then changes by at most $0.16\%$ for all the pairs we tested, up to $u=30$ with $u\equiv p/(k\,e^{\Nw})$, and the same bound holds for every closure when the pair freezes before or during the release ($u\le3$). 
For $c_s^2<w$ the post-release tail is numerically ill-conditioned in our formulation (see Appendix~\ref{app: Uniform density versus zero shear}).

We also tested a non-passive sink, adding to the radiation either a non-adiabatic pressure $\delta p_{\rm nad}=\lambda(\rho+p)\mathcal B_V$ or a scalar anisotropic stress $\Pi_e=\lambda(\rho+p)\mathcal B_K$, active while the vacuum depletes and through the exit. Here, $\mathcal B_V$ is the local volume bilinear of the graviton pair and $\mathcal B_K$ is a kinetic bilinear that vanishes once the internal modes freeze (see more details in Appendix~\ref{app: Uniform density versus zero shear}). 
We find that the non-adiabatic pressure changes the post-exit $\zeta$ by $0.011\lambda$, a local term of order $\chi\chi$ whose share of the enhanced curvature falls as $x^{-2}$ ($2\times10^{-3}$ at $x=3$ to $2\times10^{-6}$ at $x=100$). 
The anisotropic stress, entering the traceless constraint through $\nabla^{-2}$, carries the same $(p/k)^2$ as the gravitons' own stress. For $\lambda=1$ it lowers $|\zeta|$ by $4.8$--$4.9\%$ for all $x$, and by $4.3\%$ for $\epsilon=0.003$. The change is linear in $\lambda$; it rescales $c_w$ and moves the window by $0.1\lambda$ e-folds. 
A persistent stress $\propto\chi\chi$, as in a purely elastic medium, would instead grow as $(aH/k)^2$ and dominate the spectrum, which excludes a response of that form. 
A viscoelastic response, whose memory of the strain relaxes on a finite time, as well as and a spatially non-local response lie outside these tests. 
None of the responses we tested selects pairs. A response that is local in space cannot distinguish $p\lesssim k\,e^{\Nw}$ from $p\gtrsim k\,e^{\Nw}$ without correlations over $e^{\Nw}$ Hubble radii. A produced sector that implemented the window would carry exactly the non-locality that we assign to the clock, since it would be a realisation of the clock, with its density, entropy, stress and memory entering its dynamics. Also, a response of the sink correlated with the gravitons does not double count their stress, because $Q^\mu{}_\nu$ is the Bunch-Davies fluctuation, with no occupied quanta, while the sink is made of the quanta that the depletion produces.
Within the passive closure, then, the sink adds no free function to the observables: its enthalpy is fixed by the tilt (Sec.~\ref{sec: Spectrum and tilt}), and the other responses we tested either drop out or rescale $c_w$. Clearly, these tests cover the responses listed here, not every medium the depletion could produce.

The window is distinct from the end of dS. The condition $p\lesssim k\,e^{\Nw}$ is a fixed ratio between internal and external scales, i.e. $x_w=e^{\Nw}$, whereas the requirement that the internal modes freeze before the end, $x_{\rm end}=\kend/k$, depends on $k$. Notice that a fixed ratio is the prescription of Ref.~\cite{Traforetti:2025cax}, which takes $x_{\max}=e^{N_{\rm obs}}=\kend/H_0$ for every $k$. 
Our clock realises that prescription dynamically and adds the transfer to the post-exit curvature. However, $x_{\max}=e^{N_{\rm obs}}$ needs internal modes up to $k\,e^{N_{\rm obs}}>\kend$ for every $k>H_0$, and these modes never leave the horizon before the end. With both conditions, the effective count is $x_{\rm eff}=\min(e^{\Nw},\kend/k)$, and it is the same for every scale only below $k_{\rm b}=\kend e^{-\Nw}$, beyond which it produces the $k^{-1}$ tail (see Sec.~\ref{sec: Spectrum and tilt}). Hence, $\Nw$ has to be shorter than $N_{\rm obs}$ by at least $\ln(k/H_0)$, plus half the release width (Sec.~\ref{sec: Observational consequences}). For the smallest scale that must be scale invariant, this implies about 5 e-folds at the pivot and 8 at $k=1\,$Mpc$^{-1}$. For $\Nw=N_{\rm obs}$ the break would sit at $k_{\rm b}=H_0$.
In our realisation, the clock selects the pairs through its perturbation $\delta\rho_v(\mathbf k)=\bar\rho_v'\,\xi(\mathbf k)$, an isotropic spin-0 response whose pair kernel sets the ratio $|\zeta|/|\zeta_{\rm P}|$ of Fig.~\ref{fig:window}. This relation identifies the clock slices, but does not give the microscopic equation for $\xi$, which may involve the produced sector and the vacuum. We therefore state the selection as a requirement on any microscopic completion: the response of the coupled system of depleting vacuum and produced sector has to reproduce the fixed ratio $p/k\lesssim e^{\Nw}$, and the density, entropy, stress and memory of the produced medium may all take part in it. 
The window makes the spectrum scale invariant, and the tilt comes from the drift of $H$ across it.

\subsection{Solving through the transition}
\label{sec: Solving through the transition}
We integrate the exact second-order equations for single graviton pairs from inside the horizon, at $k|\eta|=5$ with Bunch-Davies modes, through the dS phase and a smooth exit, and into the radiation era. The background is quasi-dS, with a vacuum energy $\rho_v\propto e^{-2\epsilon N}$ that decays slowly into the radiation sink of Eq.~\eqref{eq:sinkbg} with $\epsilon=0.0088$, and at the end converts completely over $\Delta N=0.3$ e-folds on the clock slices. We evolve the scalar sector in Poisson gauge, Eqs.~\eqref{eq:00}--\eqref{eq:zz}, with the vacuum perturbation above and the clock of Eq.~\eqref{eq:clock}, while the gravitons evolve on the exact background. After the exit we evaluate $\zeta$ on uniform-density slices, Eq.~\eqref{eq:zetaUD}, which for radiation is $\zeta=-\tfrac12\phi-\tfrac1{12}\chi\chi+\tfrac14\delta\rho_r/\rho_r$. The details are in Appendix~\ref{app: Numerical scheme for the window clock}.

A clock that is never released reproduces a direct Poisson-gauge evolution to $10^{-11}$. A clock released deep inside the horizon approaches the proper-time value, $|\zeta|=0.125$ for release at $p|\eta|=546$ with $x=10$, compared with $|\zeta|=0.129$ from a synchronous-gauge evolution on the same background. Moving the exit from 3 to 10 e-folds after the release changes $\zeta$ by less than $10^{-6}$.

\begin{figure*}[t]\centering
\includegraphics[width=\textwidth]{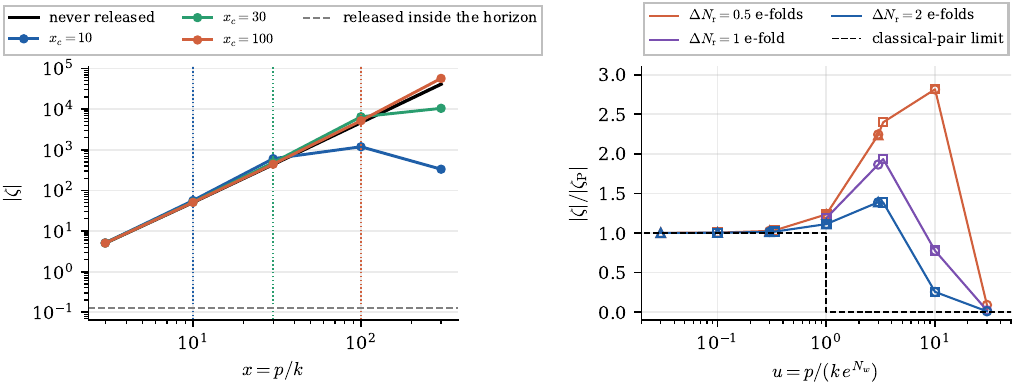}
\caption{\textit{Left}: Curvature $|\zeta|$ after the exit for one pair ($y=x+0.4$, positive helicities), with windows $x_c=e^{\Nw}=10$, $30$ and $100$ released over $\Delta N_{\rm r}=2$ e-folds. Each curve tracks the never-released zero-shear value (solid black line) up to its window $x\simeq x_c$ (dotted lines) and falls steeply beyond. The gray dashed line is a clock released inside the horizon. \textit{Right}: Ratio $|\zeta|/|\zeta_{\rm P}|$ against $u=x/x_c=p/(ke^{\Nw})$, for three release widths. Curves with different $x_c$ at the same $u$ agree to better than half a per cent. The dashed black line is the schematic classical-pair limit, in which only pairs frozen before the release are recorded.}
\label{fig:window}
\end{figure*}

Figure~\ref{fig:window} shows the result for the curvature $|\zeta|$. We see that for $u\ll1$ the curvature is the zero-shear one, while for $u\gg1$ it falls steeply, as $u^{-3}$ between $u=10$ and $u=30$ for $\Delta N_{\rm r}=2$. At fixed $u$, the ratio depends on $x_c$: at $u=10$ and $\Delta N_{\rm r}=2$, for example, $(x,x_c)=(100,10)$ and $(300,30)$ give $0.253$ and $0.254$. Near $u\sim1$--$10$, the kernel is enhanced, because pairs that cross the horizon during the release hand their partly formed offset to the clock. The effect shrinks as the release slows down. At $u=10$ the $|\zeta|/|\zeta_{\rm P}|$ is $2.8$, $0.77$ and $0.25$ for $\Delta N_{\rm r}=0.5$, $1$ and $2$, while at $u=30$ we get $|\zeta|/|\zeta_{\rm P}|=0.09$, $0.02$ and $0.008$, respectively.

\subsection{Spectrum and tilt}
\label{sec: Spectrum and tilt}
Since the kernel is the same for every $k$,
\begin{equation}
\Delta^2_\zeta(k)=\frac{c_w}{20}\,\Delta_h^4\big(k\,e^{\Nw}\big)\,e^{\Nw}\,,
\label{eq:spectrum}
\end{equation}
with
\begin{equation}
c_w\equiv\frac5{16}\,e^{-\Nw}\int_0^\infty g^{\rm rec}_\zeta\,\dd x \,,
\label{eq:cw}
\end{equation}
where $g^{\rm rec}_\zeta$ is Eq.~\eqref{eq:loop} evaluated with the curvature that each pair leaves after the exit, so that $g^{\rm rec}_\zeta\to g_\zeta f^2$ for $x\gg1$. The smallest value, $c_w=1.17$, is obtained from the full loop over all configurations and helicities for a sharp release, in which each pair is recorded with the zero-shear value it has when the window closes. For pairs that freeze close to the release, this exceeds their late-time value. A release spread over a finite number of e-folds also records part of the offset of pairs that cross the horizon during the release, and from the positive-helicity slice of the dynamical runs we estimate $c_w\approx7.8$, $18$ and $71$ for $\Delta N_{\rm r}=2$, $1$ and $0.5$, respectively. A release much faster than an e-fold would also record the sub-horizon, vacuum part of the zero-shear offset, so we regard $\Delta N_{\rm r}\gtrsim1$ as the physical range. Only the combination $c_we^{\Nw}$ enters the amplitude, so the release profile moves the lower edge of the allowed window by only $\ln(c_w/1.17)\lesssim4$.

The dominant pairs have $p\simeq(1$--$10)\,k\,e^{\Nw}$, a factor that is absorbed into $c_w$. They freeze a fixed number of e-folds after $k$ exits, with $\Delta_h^2\propto H^2\propto p^{-2\epsilon}$, and since the scalar carries four tensor legs, $\Delta^2_\zeta\propto H^4(p)$, giving the consistency relation 
\begin{eqnarray}
    n_s-1=2n_T
    \label{eq: consistency relation}
\end{eqnarray}
Then, the scalar and tensor tilts read
\begin{equation}
n_s-1=-4\epsilon,\qquad n_T=-2\epsilon,
\label{eq:tilt}
\end{equation}
respectively. The running $\alpha_s=-4\,\dd\epsilon/\dd\ln k=\mathcal O(\epsilon^2)$ is found to be $\alpha_s=+8\epsilon^2\simeq6\times10^{-4}$ if $\epsilon$ evolves as in Ref.~\cite{Jimenez:2026rqn}. The observed $n_s=0.965$~\cite{Planck:2018jri} fixes $\epsilon=0.0088$.
Since $\epsilon=\tfrac32\Omega_e(1+w_e)$, the tilt fixes the enthalpy fraction of the sink, $\Omega_e(1+w_e)\simeq(1-n_s)/6$, and not $w_e$ or $c_s^2$ separately. For example, for radiation we obtain $\Omega_r\simeq0.0044$. The relation $n_s-1=2n_T$ also assumes that the dimensionless response does not depend on scale. Indeed, writing $\Delta^2_\zeta=\Delta_h^4(k)\,\mathcal I(k)$, a scale-dependent response gives $n_s-1=2n_T+\dd\ln\mathcal I/\dd\ln k$.
The window sets the enhancement and makes it scale-independent, whereas a count running to the end for every scale would give $n_s-1\simeq-1$. Therefore, the red tilt comes entirely from the drift of $H$ across the window. Modes that exit less than $\Nw$ e-folds before the end have fewer pairs available, $x_{\rm eff}=\kend/k$, so beyond $k_{\rm b}=\kend e^{-\Nw}$ the spectrum falls as $k^{-1}$.

Because all patches end on the same clock, the radiation era starts with a single adiabatic mode and no isocurvature. The curvature is a sum over $\sim e^{\Nw}$ independent pairs, and power counting of the one-loop bispectrum gives $f_{\rm NL}\sim\Nw/(\Delta_h^2e^{2\Nw})\lesssim10^{-14}$ at the lower edge of the window, the same scaling as in Ref.~\cite{Abdelaziz:2026qyk}.

\subsection{Conservation through the radiation era}
\label{sec: Conservation through the radiation era}
\begin{figure}[t]\centering
\includegraphics[width=\columnwidth]{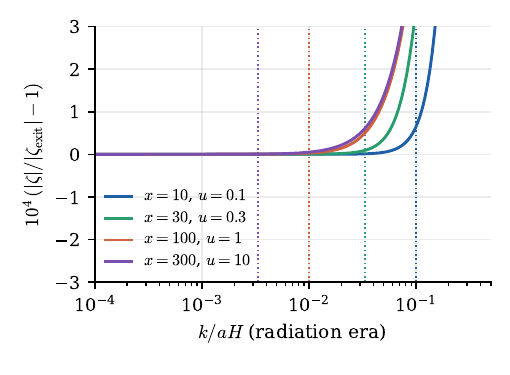}
\caption{Relative change of $\zeta$ after the exit, $10^4(|\zeta|/|\zeta_{\rm exit}|-1)$, against $k/aH$ in the radiation era, for four pairs inside ($u=0.1,0.3,1$) and outside ($u=10$) the window. Dotted vertical lines mark the re-entry of the internal modes, $p=aH$. On super-horizon scales, the curvature $\zeta$ is conserved to better than $10^{-4}$.}
\label{fig:cons}
\end{figure}

After the exit, the matter is a single radiation fluid. The frozen gravitons stay frozen until they re-enter, and then behave as radiation with a locally determined stress. The non-adiabatic pressure vanishes, so $\zeta$ on uniform density is conserved on super-horizon scales~\cite{Wands:2000dp,Weinberg:2003ur}. Figure~\ref{fig:cons} shows this for pairs inside ($u=0.1,0.3,1$) and outside ($u=10$) the window.
From the exit until the external mode approaches the horizon, $\zeta$ stays constant between $1\times10^{-6}$ and $6\times10^{-5}$ while $k/aH<1/30$, including across the re-entry of the internal modes. The rise near $k/aH\sim0.1$ is the usual $(k/aH)^2$ gradient term of the adiabatic mode, and after that $\zeta$ sets the amplitude of the acoustic oscillations in the standard way.

\section{Observational consequences}
\label{sec: Observational consequences}
\subsection{The allowed window}
\label{sec: The allowed window}
Several numbers of e-folds enter here, and they have to be kept apart. The duration of dS, $N_{\rm tot}$, runs from the exit of the first mode to the end and is set by the lifetime of the depleting state \footnote{In Ref.~\cite{Jimenez:2026rqn}, $N_{\rm tot}\simeq100$--$114$ e-folds at saturation for $C_{\rm q}=1$.}. Here, $N_{\rm tot}$ only has to exceed $N_{\rm obs}$. In Ref.~\cite{Traforetti:2025cax}, the observable number of e-folds is taken to be $N_{\rm obs}=\ln(\kend/H_0)$, the separation between the largest observable scale and the last mode to exit. $N(k)=\ln(\kend/k)$, with $N_*=N(k_*)$, is the time that scale $k$ spends outside the horizon before the end. $\Nw$ is the part of that time over which $k$ collects pairs, and it sets the amplitude through $e^{\Nw}$. Finally, $\Delta N_{\rm r}$ and $\Delta N=0.3$ are the widths of the release and of the exit. Only $\Nw$ is a parameter of the clock. Because $N(k)=\ln(\kend/k)$ depends only on the comoving scales, it is fixed by $H_{\rm end}$ and by what follows the end of dS, and lengthening the dS phase raises $N_{\rm tot}$, but leaves $N(k)$ unchanged.

We take IWI's natural exit, directly into radiation at $\rho_{\rm end}=3M^2H_{\rm end}^2$ with $M$ the reduced Planck mass. The number of e-folds between the pivot $k_*=0.05\,$Mpc$^{-1}$ and the end is then fixed by $H_{\rm end}$. Because of the drift, the Hubble rate at the end is lower than at the pivot, $H_{\rm end}=H_*e^{-\epsilon N_*}$, and the legs that set the amplitude at the pivot freeze with $H=H_*e^{-\epsilon\Nw}$. For each $\Nw$, Eq.~\eqref{eq:spectrum} with $A_s=2.1\times10^{-9}$ then fixes $H_*$, the tensor-to-scalar ratio $r=\Delta_h^2(k_*)/A_s$ and $N_*$. Since $r$ falls as $\Nw$ grows, the window is bounded from below by the BICEP/Keck limit $r_{0.05}<0.036$~\cite{BICEP:2021xfz}, where $k_*=0.05\;\rm Mpc^{-1}$.

From above, every scale that has to be scale invariant needs its window to close before the end. The release of scale $k$ is complete $\Nw+\Delta N_{\rm r}/2$ e-folds after it exits. With the exit placed one e-fold after the centre of the release, for $\Delta N_{\rm r}=2$ and $1$, the post-exit curvature of pairs with $u=0.3$--$30$ agrees with that of a late exit to $0.2\%$, and $\int f^2\,\dd u$ to $0.3\%$, although the pairs with $u\gtrsim3$ are still inside the horizon at the exit. After the release, the clock reads local proper time, so the offset it has recorded is complete and the remaining pairs do not have to freeze before the end. An exit during the release instead records more of the zero-shear offset, $f(3)=2.08$ instead of $1.39$ for an exit at the centre of a release with $\Delta N_{\rm r}=2$, so the scales whose release is under way at the end form a transition of width $\Delta N_{\rm r}$ around $k_{\rm b}$ before the $k^{-1}$ tail. Requiring scale invariance down to $k=1\,$Mpc$^{-1}$ therefore means $\Nw+\Delta N_{\rm r}/2\le N(1\,{\rm Mpc}^{-1})$. For the sharp release, $c_w=1.17$, this gives
\begin{equation}
30.5\le\Nw\le50.3,\qquad2.6\times10^{-6}\le r\le0.036\,.
\label{eq:window}
\end{equation}
For $c_w=7.8$ ($\Delta N_{\rm r}=2$) the range is $28.6\le\Nw\le49.2$. Across the window $N_{\rm obs}=N_*+5.4$ runs from $61.1$ to $58.7$, so $\Nw$ is between 8 and 31 e-folds shorter than $N_{\rm obs}$.

These bounds hold for the direct radiation exit. Under this exit, the literal prescription of Ref.~\cite{Traforetti:2025cax}, $\Nw=N_{\rm obs}$, has no room, because for every $k>H_0$ it would need pairs that exit after the end. A history with a phase of different equation of state between the end of dS and the radiation era changes $N(k)$, and with it the upper bound. That would be a different realisation, which needs its own curvature matching, so we do not consider it here. 
Table~\ref{tab:window} lists the results for a direct radiation exit. The reheating temperature is $T_{\rm rh}\simeq(0.4$--$4.5)\times10^{15}\,$GeV, and the end of dS corresponds to $f_{\rm end}\simeq(1.1$--$12)\times10^{7}\,$Hz today.

\begin{table}[t]\centering\small
\begin{tabular}{rccccc}\toprule
$\Nw$ & $r$ & $H_*$ [GeV] & $N_*$ & $k_{\rm b}$ [Mpc$^{-1}$] & $10^8\mu$ \\\midrule
30.5 & $3.6\times10^{-2}$ & $4.7\times10^{13}$ & 55.7 & $4.2\times10^{9}$ & 1.65\\
32   & $1.8\times10^{-2}$ & $3.3\times10^{13}$ & 55.5 & $8.3\times10^{8}$ & 1.65\\
35   & $4.2\times10^{-3}$ & $1.6\times10^{13}$ & 55.2 & $2.9\times10^{7}$ & 1.65\\
40   & $3.8\times10^{-4}$ & $4.8\times10^{12}$ & 54.6 & $1.1\times10^{5}$ & 1.65\\
43   & $8.8\times10^{-5}$ & $2.3\times10^{12}$ & 54.2 & $3.7\times10^{3}$ & 1.61\\
45   & $3.4\times10^{-5}$ & $1.4\times10^{12}$ & 54.0 & $3.9\times10^{2}$ & 1.23\\
48   & $7.9\times10^{-6}$ & $7.0\times10^{11}$ & 53.6 & $14$ & 0.24\\
50.3 & $2.6\times10^{-6}$ & $4.0\times10^{11}$ & 53.3 & $1.0$ & 0.02\\\bottomrule
\end{tabular}
\caption{Direct radiation exit with the Hubble drift included, for $c_w=1.17$, $A_s=2.1\times10^{-9}$ and $n_s=0.965$. We report: the window $\Nw$; the tensor-to-scalar ratio $r$ at the pivot $k_*=0.05\,$Mpc$^{-1}$; the Hubble rate $H_*$ at $k_*$; the e-folds from the pivot to the end, $N_*=\ln(\kend/k_*)$; the small-scale break $k_{\rm b}=\kend e^{-\Nw}$; and the chemical-potential distortion $\mu$ for a sharp break. Without the break, $\mu=1.65\times10^{-8}$ throughout.}
\label{tab:window}
\end{table}

\begin{figure*}[t]\centering
\includegraphics[width=\textwidth]{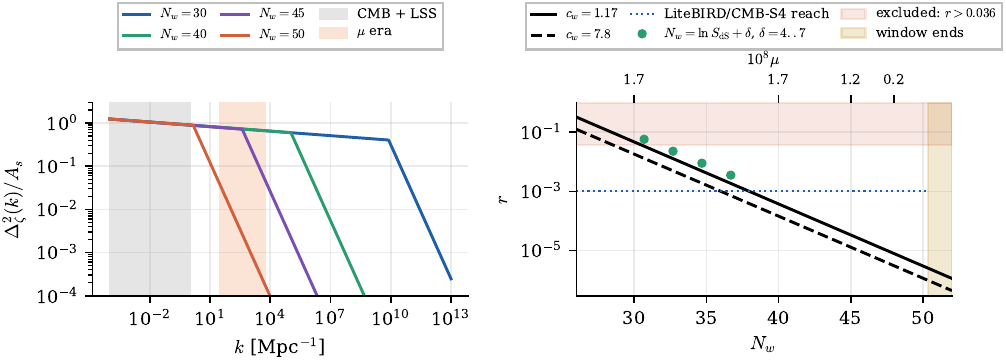}
\caption{\textit{Left}: Scalar spectrum $\Delta^2_\zeta(k)/A_s$ for $\Nw=30,40,45,50$. The spectrum is flat, with $n_s-1=-4\epsilon$, through the CMB and large-scale-structure range (gray band), and steepens to $k^{-1}$ beyond $k_{\rm b}$. The orange band marks the $\mu$-distortion scales. \textit{Right}: Tensor-to-scalar ratio $r$ against $\Nw$ for $c_w=1.17$ (solid line) and $c_w=7.8$ (dashed line). The upper shaded region is excluded because $r>0.036$, while the right-hand one is excluded because the window would reach the end of de Sitter for $k<1\,$Mpc$^{-1}$. The dotted line is the LiteBIRD/CMB-S4 reach. Circles are the scrambling-time windows of Eq.~\eqref{eq:scramble}. The top axis gives $\mu$-distortion for $c_w=1.17$.}
\label{fig:spec}
\end{figure*}

\subsection{Tensor-to-scalar ratio and consistency relation}
\label{sec: Tensor-to-scalar ratio and consistency relation}
The tensor-to-scalar $r$ falls by a factor $e^{-(1-4\epsilon)/2}$ per unit of $\Nw$ (Fig.~\ref{fig:spec}, right). B-mode programmes that target $r\sim10^{-3}$~\cite{LiteBIRD:2022cnt,CMB-S4:2016ple} probe $\Nw\lesssim38$ for $c_w=1.17$, or $\Nw\lesssim36$ for $c_w=7.8$. Whatever the value of $r$, the tensor tilt in IWI is $n_T=(n_s-1)/2\simeq-0.018$, which differs from the single-field relation\footnote{The single-field consistency relation $n_T=-r/8$ would give $-0.001$ for $r=10^{-2}$.} $n_T=-r/8$.

A natural candidate for the window is the dS scrambling time, $t_*=H^{-1}\ln S_{\rm dS}$, with $S_{\rm dS}=\pi\mpl^2/H^2$, which is the time over which the degrees of freedom of a horizon scramble information~\cite{Sekino:2008he,Susskind:2011ap,Aalsma:2020aib}. Identifying it with the window assumes that regions which have left each other's horizon stay coherent for that long. Setting $\Nw=\ln S_{\rm dS}+\delta$, evaluated when the window closes, gives $\Delta_h^2e^{\Nw}=16e^{\delta}$ exactly and
\begin{equation}
r=\frac{5}{4\,c_w}\,e^{-\delta}\,e^{2\epsilon\Nw},\qquad n_s-1=n_T=-2\epsilon,
\label{eq:scramble}
\end{equation}
where the factor $e^{2\epsilon\Nw}$ accounts for the drift between the pivot and its legs. The ratio then no longer depends on $H$. The tilt changes as well, because $e^{\Nw}\propto H^{-2}$ when the window closes. With $\epsilon=0.0088$ the bound $r<0.036$ requires $\delta\ge4.5$ for $c_w=1.17$ and $\delta\ge2.5$ for $c_w=7.8$, so the window has to exceed the scrambling time by a few e-folds. For $\delta$ up to $7$, the window is $\Nw\simeq30$--$37$ for $c_w=1.17$ and $r$ lies between $3\times10^{-3}$ and the current limit, within reach of the next generation of B-mode experiments. Such a window would violate the consistency relation $n_s-1=2n_T$ of Eq.~\eqref{eq: consistency relation} and Ref.~\cite{Jimenez:2026rqn}, and it needs a drift twice as large. A measurement of $n_T$ therefore separates the two cases.

\subsection{The gravitational-wave background today}
\label{sec: The gravitational-wave background today}
The stochastic background is dominated by the first-order tensors. For modes that re-enter during radiation domination~\cite{Watanabe:2006qe}, the stochastic gravitational-wave background today is
\begin{equation}
\Omega_{\rm GW}h^2(f)=\frac{\Omega_rh^2}{24}\,\frac{g_*(T_f)}{g_{*0}}\Big(\frac{g_{s0}}{g_s(T_f)}\Big)^{4/3}\Delta_h^2(k_*)
\Big(\frac{f}{f_*}\Big)^{n_T},
\label{eq:omega}
\end{equation}
up to $f_{\rm end}$, with $\Omega_rh^2=4.18\times10^{-5}$ and $T_f$ the temperature at re-entry.

\begin{figure*}[t]\centering
\includegraphics[width=\textwidth]{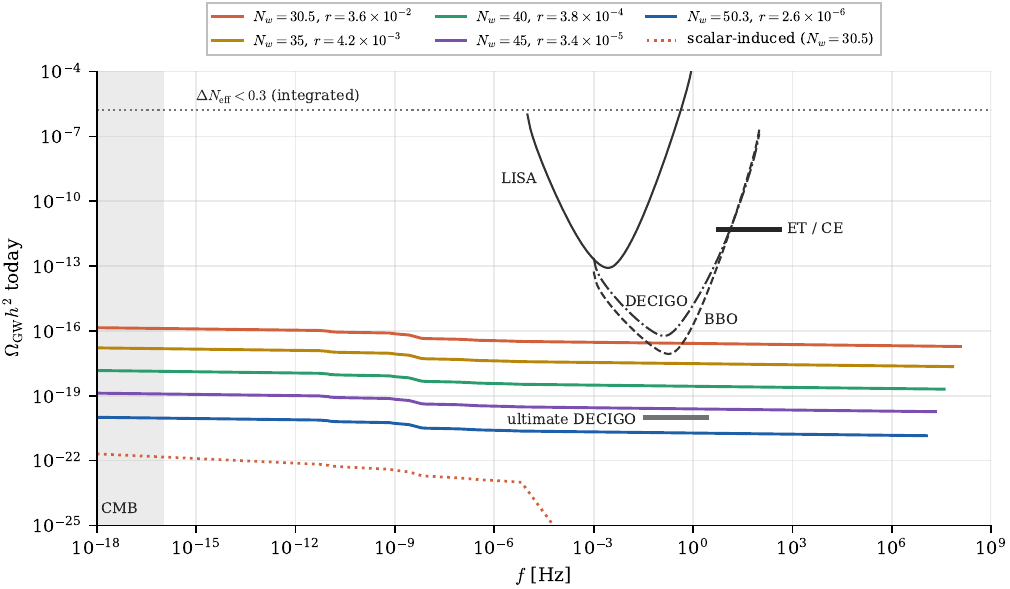}
\caption{Gravitational-wave background today, $\Omega_{\rm GW}h^2$, against frequency $f$, across the allowed window (solid lines, labelled by $\Nw$ and $r$), with the scalar-induced background for $\Nw=30.5$ (dotted red line). Black curves are approximate power-law-integrated sensitivities (SNR $=10$, four years) for LISA, BBO and DECIGO. Bars mark representative levels for an ultimate DECIGO and for ET/CE. The dotted gray line is the integrated $\Delta N_{\rm eff}<0.3$ bound and the gray band the CMB scales. Each spectrum ends at $f_{\rm end}$, while the steps are the changes in $g_*$.}
\label{fig:gw}
\end{figure*}

Figure~\ref{fig:gw} shows the result for $\Omega_{\rm GW}h^2(f)$ across the window. At $0.3\,$Hz, $\Omega_{\rm GW}h^2$ runs from $2.7\times10^{-17}$ at the lower edge to $1.9\times10^{-21}$ at the upper edge. We compare it with approximate power-law-integrated sensitivities~\cite{Thrane:2013oya} for four years of data at signal-to-noise ratio 10 (see more details in Appendix~\ref{app: Gravitational-wave spectrum and sensitivities}), assuming that astrophysical foregrounds can be subtracted. BBO~\cite{Crowder:2005nr} detects the background for $r\gtrsim0.011$, that is $\Nw\lesssim33$ for $c_w=1.17$. The baseline DECIGO~\cite{2011CQGra..28i4011K} would need $r\gtrsim0.08$, which is already excluded, while LISA~\cite{LISA:2017pwj} and ground-based interferometers~\cite{Punturo:2010zz} are far from the signal. An ultimate DECIGO, with $\Omega h^2\sim10^{-20}$~\cite{Kudoh:2005as}, reaches $r\gtrsim1.3\times10^{-5}$, or $\Nw\lesssim47$. Expressed in terms of $r$, these reaches depend neither on the normalisation convention nor on $c_w$.

The scalar-induced gravitational wave background is $0.82\,A_s^2\,\Omega_rh^2$ times the $g_*$ factor of Eq.~\eqref{eq:omega}, about $6\times10^{-23}$ on scales $k<k_{\rm b}$, and it falls steeply beyond. It was computed at second order in the Einstein-de Sitter and radiation eras in Refs.~\cite{Matarrese:1997ay,Mollerach:2003nq} and revisited in Refs.~\cite{Ananda:2006af,Baumann:2007zm,Kohri:2018awv}. Tensor-induced tensors are of order $\Delta_h^4\Omega_r\lesssim10^{-25}$. The integrated energy density is at most $\int\Omega_{\rm GW}h^2\,\dd\ln f\simeq3\times10^{-15}$, reached at the lower edge, nine orders of magnitude below the $\Delta N_{\rm eff}$ bound. Therefore, both contributions are negligible throughout the window. 

\subsection{Spectral distortions and small scales}
\label{sec: Spectral distortions and small scales}
The break at $k_{\rm b}$ is a distinctive feature of the mechanism. For $\Nw\lesssim42$ it lies beyond the scales that produce $\mu$-distortions, and $\mu$ keeps its standard value, $1.65\times10^{-8}$ for these parameters~\cite{Chluba:2012we}. For larger windows the break moves into the $\mu$ era, $31.6\lesssim k/{\rm Mpc}^{-1}\lesssim5400$, and $\mu$ drops to $1.6\times10^{-8}$ at $\Nw=43$, $1.2\times10^{-8}$ at $\Nw=45$, $2.4\times10^{-9}$ at $\Nw=48$ and $2\times10^{-10}$ at the upper edge of the window. A measured $\mu$ below the standard value, with no detected B-modes, would therefore point to a long window. The same steepening suppresses the small-scale power relevant for primordial black holes and for the induced background at high frequencies, as these values assume a sharp break. For a release of finite width, the scales whose release is under way at the end receive more than the plateau value (Sec.~\ref{sec: Observational consequences}), so the spectrum can rise before it falls; we have not computed the shape of this transition, and for $\Nw\gtrsim43$ the value of $\mu$ is uncertain in both directions.

\section{Discussion}
\label{sec: Discussion}
We have shown that the tensor modes of  Inflation without an Inflaton (IWI) are not enhanced at second order, and that the enhanced graviton-induced curvature can be carried into the radiation era with the observed tilt.

The tensors are not enhanced because the transerve-traceless (TT) sector obeys a wave equation whose retarded Bunch-Davies response is suppressed by oscillatory cancellations across the full time integral, whereas the scalar potential is fixed by an elliptic constraint that responds at once to the stress of all frozen gravitons. The gravitational waves of IWI are then the vacuum gravitons, Gaussian to a part in $10^{5}$, with tesnro tilt $n_T=-2\epsilon$, as long as the depletion adds no long-range spin-2 response of its own (Sec.~\ref{sec: Momentum window versus temporal support}).

For the scalar, the curvature that survives is that of the hypersurface on which de Sitter (dS) ends, and its enhanced part is the offset between this hypersurface and local proper time, which the clock carries by following zero-shear time. In the $\delta N$ picture, the inflaton is both the source of this offset and the clock that reads it. In IWI the two roles can be distinguished, although their dynamics need not be independent. The Bunch-Davies gravitons provide the source through the second-order constraints, and in our construction the depleting vacuum provides a clock that reads it within a finite reach, $p/k\lesssim e^{\Nw}$, which makes the enhancement scale independent.
This is the fixed-ratio prescription of Ref.~\cite{Traforetti:2025cax}, realised in time and made compatible with the end of dS. The non-vacuum sector fed by the depletion is the sink required by energy-momentum conservation, i.e. the counterpart of the reheating products. For adiabatic sinks, we find that it changes the amplitude by less than $0.2$ per cent. The radiation era then begins with a single adiabatic and Gaussian mode, conserved on super-horizon scales, with amplitude $c_w\Delta_h^4e^{\Nw}/20$ and tilt $n_s-1=-4\epsilon$, which steepens beyond $\kend e^{-\Nw}$. On the other hand, Refs.~\cite{Jimenez:2026rqn,bridge} read the spectrum at horizon crossing, where the populated pairs have loop momenta of order the external scale. The result for the scalar tilt $n_s-1=-4\epsilon$ is the same, but which of the two readings carries the observed amplitude is to be investigated.

Several results follow directly from the Einstein equations: the sourcing of the scalar by graviton pairs, the $(p/k)^2$ enhancement of the zero-shear curvature, the consistency of the pure-$\Lambda$ solution with all four scalar second-order equations, the absence of tensor enhancement in the hard-region loop with Bunch-Davies sources and the need for a sink once the vacuum depletes. We have checked numerically that the post-exit uniform-density curvature is conserved for the radiation sink, and that it is insensitive to the sink for adiabatic sinks and for $c_s^2>w$. However, for $c_s^2<w$ our runs did not converge, so we do not draw any conclusion from them. The local entropy and anisotropic responses we tested either leave $\zeta$ unchanged up to an unenhanced local term or rescale $c_w$ by about $5\lambda$ per cent for a response of relative strength $\lambda$, while a persistent stress proportional to the strain is excluded. We assume the global zero-shear clock, which we identify with the break of the dS coherent state as a whole and which synchronises regions up to $10^{13}$--$10^{22}$ Hubble radii apart. We also assume the value of the window, with a few e-folds beyond the scrambling time as a natural candidate; the memory of the clock after the release, which keeps the offset it has acquired instead of resetting to local proper time; the release profile, which sets $c_w$ but moves the lower edge of the window by only a few e-folds; and the end of dS, which a steady depletion does not produce. The transfer of the enhanced curvature into the radiation era therefore follows, given the hypotheses of the adopted clock: synchronisation on zero-shear time for $\Nw$ e-folds, release, and memory afterwards. How the system composed of the vacuum and produced sector realises them remains open to further investigation.

Moreover, within the passive closure, we have seen that the equation of state of the sink matters little. The main open problem is to derive from the dynamics of the coherent state how the depletion synchronises each scale with zero-shear time for a fixed $\Nw$, how it realises the finite reach $p/k\lesssim e^{\Nw}$ and why it keeps its memory afterwards, and it has to do so through the response of the vacuum and the produced sector together. The same calculation should fix the transfer $J_\nu$ and the perturbative properties of the sink, including any density, entropy, stress and memory response, decide how the produced sector takes part in the selection, and establish whether the depletion produces a long-range spin-2 response, which would be a new source of primordial gravitational waves. We are studying these questions in ongoing work with G.~Dvali on horizon memory and a state-resolved quantum-break clock~\cite{DvaliMemory,DvaliClock}, which separates the coherent occupation of the dS state from the activity of its microscopic channels. Its connection to the $p/k$ selection remains to be established.

A natural candidate for the microscopic identity of the clock is suggested by the operator-algebraic program of Gómez \cite{Gomez:2022eui,Gomez:2023wrq,Gomez:2023upk,Gomez:2024kuy}. There, a clock is not optional: the local observables of a dS patch form a type $III_1$ factor whose reparametrization-invariant subalgebra is trivial, so gauge-invariant observables exist only after the algebra is extended by a clock and the operators are clock dressed (the crossed product), and the $\epsilon=0$ limit of the gauge-invariant algebra is argued to be algebraically nonexistent \cite{Gomez:2024kuy}. Gómez's model-independent clock for the planar patch \cite{Gomez:2023upk} has, one by one, the properties postulated in Sec.~\ref{sec: A zero-shear clock with a finite window}. It is a two-mode system, one mode inside and one beyond the horizon, whose state, i.e. the EPR squeezed pair generated by the de Sitter Bogolyubov transformation, is coherent across super-horizon distances: the non-locality the window requires is the entanglement of the clock state. It is a spin-0 number-phase system, $[\hat N,\Phi]=-i$, so it adds no long-range spin-2 response (Sec.~\ref{sec: Momentum window versus temporal support}). Its energy is $\langle H_c\rangle=\epsilon H^2a^2$, which is the enthalpy of the sink, so the background bookkeeping is already the clock's. And the gravitational bound on clock precision gives the dressing a maximal reach \cite{Gomez:2023upk}, of the kind a finite window needs, with $\ln S_{\rm dS}$ its natural scale, i.e. the scrambling-time candidate of Eq.~\eqref{eq:scramble}. What the algebraic clock does not yet provide is the value of $\Nw$ and the release-and-memory law of Sec.~\ref{sec: A zero-shear clock with a finite window}; these are precisely the properties the state-resolved quantum-break clock is meant to derive in \cite{DvaliClock}. If the identification holds, the window clock postulated here is not an addition to the scenario but the crossed-product dressing that any dS observable requires.

Several predictions are independent of these details. The tensor tilt is directly connected to the scalar one via $n_T=(n_s-1)/2$, or $n_s-1$ for a scrambling-time window, whatever the value of $r$, , so a measurement of $n_T$ both tests the mechanism and distinguishes the two cases. The tensor-to-scalar $r$ ratio lies between $3\times10^{-6}$ and the current limit, and the scalar spectrum has a small-scale break whose position is tied to $r$ through $\Nw$. Finally, the gravitational-wave background is that of the vacuum gravitons, within reach of BBO for $r\gtrsim0.01$ and of an ultimate DECIGO for $r\gtrsim10^{-5}$.

\begin{acknowledgments}
The work of RJ is supported by the Simons Foundation. Funding for the work of MT and RJ was partially provided by project PID2022-141125NB-I00, and grant CEX2024-001451-M funded by MICIU/AEI/10.13039/501100011033. DB and SM acknowledge support from the COSMOS network (www.cosmosnet.it) through ASI (Italian Space Agency) Grants 2016-24-H.0, 2016-24-H.1-2018 and 2020-9-HH.0. 
This work is supported in part by the MUR Departments of Excellence grant “Quantum Frontiers” of the Physics and Astronomy Department of Padova, Italy.
MA acknowledges the hospitality of the Institute of Cosmos Sciences of the University of Barcelona (ICCUB), where this work was carried out. The project that gave rise to these results received the support of a fellowship from ``la Caixa'' Foundation (ID 100010434), awarded to MT with code LCF/BQ/DI24/12070012. \\

AI usage: Claude Code was used to assist with coding, computations, and graphic generation. The ideas, inferences, and results all originate in the (human) brains of the authors. Claude and ChatGPT were used to check citations, spelling, grammar, inconsistencies and also to enhance the flow of the text and its logic. Handwritten notes were transcribed to Latex using these tools. The original text was written by the authors, who are responsible for the results, findings, and conclusions of this article.

\end{acknowledgments}

\bibliographystyle{apsrev4-2}
\bibliography{references}

\appendix
\section{Second-order equations}
\label{app: Second-order equations}
We take $\mathbf k=k\hat z$ and denote as $Q^\mu{}_\nu$ the graviton-bilinear part of $a^2G^\mu{}_\nu$. At the order of the bilinear cross term, the scalar equations in an arbitrary gauge are as follows ($8\pi G=1$):
\begin{align}
&k^2\phi+3\Hc\phi'+3\Hc^2\psi-2\Hc k^2(B-E')+Q^0{}_0=-a^2\delta\rho,\label{eq:00}\\
&\phi'+\Hc\psi+i\hat{\mathbf k}\cdot\mathbf Q^0/k=-a^2(\rho+p)\,v,\label{eq:0i}\\
&\phi-\psi-2\big[(B-E')'+2\Hc(B-E')\big]=T,\\
&\phi''+2\Hc\phi'+\Hc\psi'+(2a''/a-\Hc^2)\psi+Q^z{}_z=a^2\delta p,\label{eq:zz}
\end{align}
with $T$ defined in Eq.~\eqref{eq:traceless}. We have checked symbolically, for a generic $a(\eta)$, that under $x^\mu\to x^\mu+\xi^\mu$ the operator transforms by exactly $-\mathcal L_\xi\bar G^\mu{}_\nu$, and that its graviton-bilinear parts are the same in every gauge.

In Poisson gauge, the dynamical equation is Eq.~\eqref{eq:zz}, with $\psi=\phi-T$ and $\delta\rho$ taken from Eq.~\eqref{eq:00}. For the window clock $\delta p=\tfrac13\delta\rho_r-\delta\rho_v$ with $\delta\rho_v=\bar\rho_v'\xi$, and the curvature on uniform total density is $$\zeta=-\tfrac12\phi-\tfrac1{12}\chi\chi+\delta\rho/[3(\rho+p)].$$ 

\section{Consistency of the pure-\texorpdfstring{$\Lambda$}{Lambda} solution}
\label{app:consistency}
In pure GR$+\Lambda$ we have $\delta\rho=v=\delta p=\Pi=0$. We take $\phi$ from the comoving Poisson equation~\eqref{eq:poisson}, and $\psi=\phi-T$ from Eq.~\eqref{eq:traceless}, with exact Bunch-Davies legs, and evaluate on this solution the full second-order operator $a^2G^\mu{}_\nu$, metric and graviton parts together. The derivatives $\phi'$, $\phi''$, $\psi'$ and $\psi''$ are those of the functions themselves. We compute them by fourth-order finite differences with step $10^{-20}|\eta|$ in 80-digit arithmetic. Table~\ref{tab:residuals} gives, for each pair, the largest residual of the four scalar projections (00, $\hat{\mathbf k}\cdot0i$, trace, and longitudinal $ij$) over seven times in $k|\eta|\in[10^{-6},5]$, each divided by the larger of its metric and graviton parts. All four vanish to the precision of the differencing, $\le5\times10^{-25}$. 
The frozen form of the same solution $\phi=\mathcal F_\chi/4$, with $\psi=\phi'=\phi''=0$, instead fails the 00, trace and longitudinal equations at order one (see Table~\ref{tab:residuals}). Its $0i$ residual is the quantity $m$ defined below, which vanishes as $k|\eta|$, but $3\Hc m$, which appears in the 00 equation, stays finite.

\begin{table}[h]\centering\small
\begin{tabular}{lcccc}\toprule
& exact & \multicolumn{3}{c}{frozen ansatz}\\
pair $(x,y)$, helicities & all four & 00 & trace & long.\\\midrule
$(10,10.4)$, $++$ & $2\times10^{-28}$ & 0.70 & 0.70 & 1.00\\
$(3,3.4)$, $+-$ & $5\times10^{-25}$ & 1.01 & 1.01 & 1.00\\
$(0.7,0.8)$, $++$ & $5\times10^{-28}$ & 0.51 & 0.51 & 1.00\\
$(0.98,0.23)$, $-+$ & $3\times10^{-28}$ & 1.28 & 1.28 & 1.00\\
$(1.05,0.15)$, $--$ & $2\times10^{-27}$ & 1.75 & 1.75 & 1.00\\\bottomrule
\end{tabular}
\caption{Relative residuals of the four scalar Einstein equations in pure GR$+\Lambda$. \textit{Exact}: the Bunch-Davies solution evaluated with its own time derivatives. We find that all four residuals are comparable, so we quote their maximum over seven times in $k|\eta|\in[10^{-6},5]$. \textit{Frozen ansatz}: $\phi=\mathcal F_\chi/4$, $\psi=\phi'=\phi''=0$, with exact Bunch-Davies legs in $Q$, at $k|\eta|=10^{-6}$. The residuals differ by equation, so we list the 00, the trace, and the longitudinal residual (the $0i$ residual vanishes as $k|\eta|$ and is omitted). The last two pairs are non-planar random configurations.}
\label{tab:residuals}\end{table}

The reason the frozen ansatz fails is that three combinations stay finite at late times although the frozen limit sets each of them to zero. Each is a power of $\Hc\propto1/|\eta|$ multiplying quantities that vanish, or tend to constants, in conformal time. 
The first is the graviton term $\Hc\chi^{ik}\chi'_{ki}$, which tends to $\chi^{ik}\nabla^2\chi_{ki}$ because $\Hc f'\to-p^2f$ for Bunch-Davies legs. It appears only in $Q^0{}_0$, since $Q^l{}_l$, $T$ and $Q^z{}_z$ have no part proportional to $\Hc$. The other two come from the 00 and the $0i$ equation.
Writing $m\equiv i\hat{\mathbf k}\cdot\mathbf Q^0/k$, which tends to $m_0\eta$, the 00 equation gives $3\Hc(\phi'+\Hc\psi)=-3\Hc m\to3m_0$. Since $2a''/a-\Hc^2=2\Hc'+\Hc^2$ in exact dS, it also gives $\phi''+2\Hc\phi'+\Hc\psi'+(2\Hc'+\Hc^2)\psi=-a^{-2}(a^2m)'\to m_0$ in the longitudinal equation. 
Keeping all three, the late-time system reads
\begin{equation}
k^2\phi=-Q^0{}_0-3m_0,\enspace\phi=T,\enspace Q^z{}_z=-m_0,\enspace Q^0{}_0=Q^l{}_l \,.
\end{equation}
The last two relations express momentum and energy conservation of the frozen graviton stress, which the Bianchi identity guarantees, while the last says that frozen tensor gradients act as a $w=-1/3$ component.

For $(x,y)=(10,10.4)$ with positive helicities, at $k|\eta|=10^{-4}$, we find $k^2\phi=107.50$ and $Q^0{}_0=-362.48$, of which $-414.64$ comes from the $\Hc\chi\chi'$ term. The $0i$ equation then gives $3m_0=254.98$, the longitudinal component is $Q^z{}_z=-84.99$ and the trace is $Q^l{}_l=-362.48$. These are real parts, and the imaginary parts are about $0.1\%$. 
If instead $\chi'=\psi=\phi'=\phi''=0$ is imposed before the limit is taken, the three finite combinations are lost: the 00 equation gives $k^2\phi=-52.2$, the trace equation $k^2\phi=-Q^l{}_l=+362.5$, the traceless equation still gives $k^2\phi=k^2T=107.5$, and the longitudinal equation demands $Q^z{}_z=0$, although it should give $Q^z{}_z=-84.99$. This is the apparent overconstraint that motivated the effective fluid of Ref.~\cite{Bertacca:2024zfb}: the second-order equations of Ref.~\cite{Bertacca:2024zfb}, including the equation for $\phi_2$, are correct, and their large-scale solution $\phi_2=\mathcal F_\chi/4$, $\psi_2=0$ is the late-time limit of the exact solution. The overconstraint is an artifact of the order in which the limits are taken.

\section{Tensor time integrals}
\label{app: Tensor time integrals}
The late-time tensor is $$h^{(2)}(\mathbf k,0)=\int_{-\infty}^0G_k(0,\eta')S(\eta')\,\dd\eta',$$ with the retarded Green function of Eq.~\eqref{eq:green} at $\eta=0$,
\begin{equation}
G_k(0,\eta')=\frac{k\eta'\cos k\eta'-\sin k\eta'}{k^3\eta'^2}\to\frac{|\eta'|}{3}\quad\text{for }k|\eta'|\ll1 .
\end{equation}
For Bunch-Davies legs, the two source structures carry $f_1f_2$ and $f_1'f_2'$, each of which is $e^{-iP\eta'}$ times a polynomial, with $P=p_1+p_2$. The integrals are then elementary once the contour is tilted into the damping direction, so that $e^{-iP\eta'}$ provides the convergence factor. The integrals carry a short-distance regulator $\delta$ (the lower limit of $|\eta'|$): the $1/\delta$ and $\ln\delta$ pieces of a cutoff regularisation cancel and leave the real closed forms of Eq.~\eqref{eq:J}. For $k\to0$, $J_{FF}\to-(P^2+2p_1p_2)/P^4$ and $J_{dd}\to2p_1^2p_2^2/P^4$, while for $p_{1,2}\gg k$ we obtain $J_{FF}\to-3/(8p^2)$ and $J_{dd}\to1/8$ (see Sec.~\ref{sec: The causal solution}).
We checked the closed forms against numerical integration along $\eta'=-s(1+0.4i)$, $s\in(0,\infty)$, finding agreement to $10^{-10}$ for $(p_1,p_2)=(0.7,0.8)$, $(3.1,2.6)$, $(20.3,20.1)$ and $(0.3,1.25)$ in units of $k$.

\section{Numerical scheme for the window clock}
\label{app: Numerical scheme for the window clock}
We use $N=\ln a$ as the time variable and evolve the vacuum and radiation densities, the two graviton modes, the Poisson-gauge potential $\phi$ with $\phi'$, and the clock offset $\xi$, with an eighth-order Runge-Kutta integrator at relative tolerance $10^{-10}$. The release is a quintic smoothstep in $\ln(\Hc/k)$ of width $\Delta N_{\rm r}$ centered on $\Nw$. While the clock is synchronised, the condition $\xi=0$ of Eq.~\eqref{eq:clock} is enforced softly, by driving $\xi$ to zero at rate $20\Hc$. The exit is a smoothstep in $N$ of width $0.3$, placed a few e-folds after both the release and the freezing of the internal pair. 
We have checked that the results are robust to where the exit is placed. Moving it to one e-fold after the centre of the release, with $x_c=e^{\Nw}=10$ and $u=x/x_c=0.3$--$30$, changes $|\zeta|$ by at most $0.2$ per cent for $\Delta N_{\rm r}=2$ and $1$ (Sec.~\ref{sec: Observational consequences}).

An unreleased clock reproduces the direct Poisson-gauge evolution to $10^{-11}$. A clock released deep inside the horizon approaches the synchronous-gauge result, $|\zeta|=0.129$ for $x=10$, releasing progressively deeper, at $p|\eta|=74$, $201$ and $546$, gives $|\zeta|=0.62$, $0.15$ and $0.125$, converging to that value from below.
The ratio to the zero-shear value depends on $x$ and $x_c$ only through $u$, to better than half a per cent. After the exit, $\zeta$ is constant to $6\times10^{-5}$ while $k/aH<1/30$ (Fig.~\ref{fig:cons}). 
These runs use a single pair configuration, $y=x+0.4$ with positive helicities, but the dilation argument of Sec.~\ref{sec: A zero-shear clock with a finite window} holds for all configurations.

\section{Uniform density versus zero shear}
\label{app: Uniform density versus zero shear}
Eliminating $\phi'+\Hc\psi$ between Eqs.~\eqref{eq:00} and \eqref{eq:0i} gives the exact decomposition
\begin{align}
\zeta&=\zeta_{\rm P}+\Hc v+\frac{\delta\rho_c}{3(\rho+p)},\label{eq:split}\\
a^2\delta\rho_c&\equiv-\Big(k^2\phi+Q^0{}_0-3i\Hc\,\hat{\mathbf k}\cdot\mathbf Q^0/k\Big)\,,\nonumber
\end{align}
where $v$ is the velocity of the sink and $\delta\rho_c$ its comoving density, which is suppressed by $(k/aH)^2$.

\begin{figure*}[t]\centering
\includegraphics{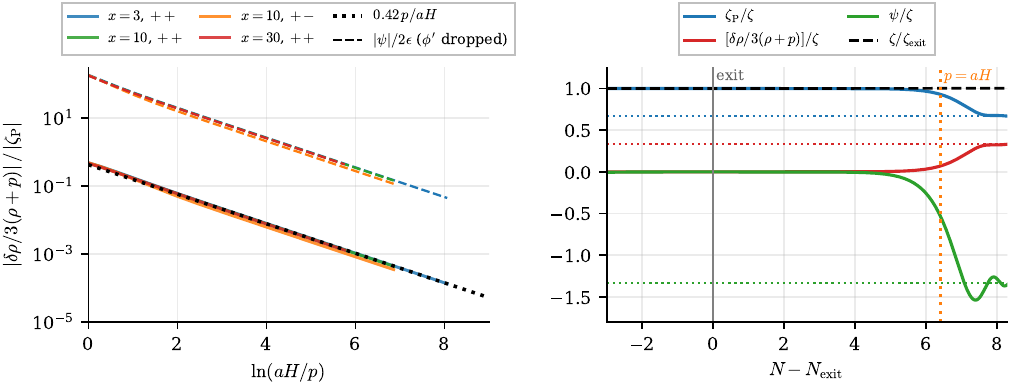}
\caption{\textit{Left}: We show the density contribution $|\delta\rho/3(\rho+p)|/|\zeta_{\rm P}|$ in quasi-de Sitter ($\epsilon=0.0088$) against $\ln(aH/p)$, the e-folds since the internal pair froze, for four pairs (solid lines overlapping). Dashed lines show the estimate $|\psi|/2\epsilon$ that dropping $\phi'$ in Eq.~\eqref{eq:00} would give. The dotted line shows $0.42\,p/aH$. \textit{Right}: We show the $x=30$ pair through a zero-shear exit and into the radiation era, against $N-N_{\rm exit}$. The split between $\zeta_{\rm P}/\zeta$ (blue) and the density term $[\delta\rho/3(\rho+p)]/\zeta$ (red) changes when the internal modes re-enter ($p=aH$, orange line), while the uniform-density $\zeta/\zeta_{\rm exit}$ (black dashed) stays constant. The lapse $\psi/\zeta$ (green) goes from $\simeq0$ while the modes are frozen to the adiabatic value as they re-enter. Dotted horizontal lines mark $2/3$, $1/3$ and $-4/3$, i.e. the adiabatic-mode values the curves relax to.}
\label{fig:density}
\end{figure*}

The left panel of Figure~\ref{fig:density} shows $|\delta\rho/3(\rho+p)|/|\zeta_{\rm P}|$ during quasi-dS for four pairs, with a zero-shear exit, $\epsilon=0.0088$ and the clock synchronised throughout. Equation~\eqref{eq:split} holds to $10^{-13}$, and the velocity term dominates, with the comoving density below $3\times10^{-3}$ of it from one e-fold after freeze-out and below $10^{-4}$ after five e-folds. A few e-folds after freeze-out, we find numerically $\Hc v=-0.1670\,\psi$ for all four pairs. This value follows from the Euler equation of the sink, which for radiation injected at rest in steady state reads $v'+4\Hc v=-\tfrac12\psi-\delta p/(\rho+p)$, the term $4\Hc v$ being the drag of the injected energy. With $\delta p/(\rho+p)=\delta\rho/[3(\rho+p)]\simeq\Hc v$ and $\psi\propto\eta$ it gives $\Hc v=-\psi/6$ up to corrections of order $\epsilon$.
Since $|\psi|\simeq(2$--$2.5)(p/aH)|\zeta_{\rm P}|$, the correction to $\zeta_{\rm P}$ is $(0.33$--$0.42)(p/aH)|\zeta_{\rm P}|$ and carries no factor $1/\epsilon$. Dropping $\phi'$ in Eq.~\eqref{eq:00} would instead have given $|\psi|/2\epsilon$, about 330 times larger (dashed curves in Figure~\ref{fig:density}, left).

After a zero-shear exit, the uniform-density curvature equals the late dS $\zeta_{\rm P}$ to $1.4\times10^{-4}$, $4.5\times10^{-4}$ and $1.3\times10^{-3}$ for $x=3$, $10$ and $30$, respectively, whose internal modes froze $8.5$, $7.3$ and $6.2$ e-folds before the exit. Both quantities approach the frozen value as $p/aH$. With the window clock released before the exit, moving the exit from 3 to 6 to 10 e-folds after the release changes $|\zeta|$ by less than $10^{-6}$ for $(x,x_c)=(10,30)$, $(3,30)$ and $(30,10)$, respectively, because the free-running clock integrates the decaying lapse.

In the radiation era, we see from the right panel of Figure~\ref{fig:density} that the split changes when the internal gravitons re-enter. While they are frozen, $\psi\simeq0$ and $\delta\rho\simeq0$. Once they re-enter, their anisotropic stress disappears, and the potentials relax to the standard adiabatic mode, $\phi\simeq\psi\simeq-\tfrac43\zeta$, with $\delta\rho/[3(\rho+p)]\to\tfrac13\zeta$ and $\zeta_{\rm P}\to\tfrac23\zeta$. The uniform-density $\zeta$ does not change, to $7\times10^{-6}$ at $k/aH=1/30$. This requires the graviton-volume term $-\chi\chi/12$: when the internal modes of the $x=3$ pair re-enter ($k/aH=1/3$) $\zeta$ drifts by $2\%$ without it and by $7\times10^{-4}$ with it.

To test the dependence on the sink, we replaced the radiation by a fluid with constant $w$ and rest-frame sound speed $c_s^2$, for which $\delta p_e=c_s^2\delta\rho_e+(c_s^2-w)\bar\rho_e'v$~\cite{Hu:1998kj}, with $\delta\rho_e=\delta\rho-\delta\rho_v$ and $\bar\rho_e'$ including the exchange. We kept $J_\nu=\partial_\nu\rho_v$ and converted the vacuum into the same fluid at the exit. The velocity term becomes $\Hc v=-\psi/(4+6c_s^2)$, independent of $w$. Table~\ref{tab:closure} gives the fractional change of $|\zeta|$ after the exit relative to radiation, for positive helicities. After a zero-shear exit, the changes of $\pm1.5\times10^{-3}$ come from the change in the background fraction of the sink, $2\epsilon/[3(1+w)]$ at fixed $\epsilon$, and the effect of $c_s^2$ is only $10^{-6}$. The window ratio $f=|\zeta_{\rm released}|/|\zeta_{\rm never\ released}|$ at $u=1/3$ lies between $1.0152$ and $1.0160$ for all closures.

When $c_s^2\neq w$ the pressure contains $(c_s^2-w)\bar\rho_e'v$, and $v$ has to be inferred from Eq.~\eqref{eq:0i} by dividing by $a^2(\rho+p)\simeq2\epsilon\Hc^2$. While the internal modes are inside the horizon, this division amplifies the integration error. For $c_s^2>w$, and for $u\le3$, the entries of Table~\ref{tab:closure} do not change when the tolerance is tightened from $10^{-10}$ to $10^{-12}$. For $c_s^2<w$ and $u\ge10$ they do change. For $(w,c_s^2)=(1,1/3)$ at $u=10$ the change is $-1.1\%$ at $10^{-12}$ and $-0.8\%$ at $10^{-13}$, and at $u=30$ it has not converged. These closures need a formulation that evolves the density and velocity of the sink explicitly.

For the responses of a non-passive sink, we kept radiation and added either $\delta p_{\rm nad}=\lambda(\rho+p)\mathcal B\,\rho_v/\rho$ to the pressure or $\Pi_e=\lambda(\rho+p)\mathcal B\,\rho_v/\rho$ to the traceless equation, $\phi-\psi=T-2a^2\Pi_e/k^2$, with the corresponding $\tfrac23a^2\Pi_e$ in Eq.~\eqref{eq:zz}. The factor $\rho_v/\rho$ switches the response off after the exit. We used the volume bilinear $\mathcal B_V=E^{(1)}:E^{(2)}f_1f_2/6$, which is the pair part of $\chi\chi/12$, and the kinetic bilinear $\mathcal B_K=E^{(1)}:E^{(2)}f_1'f_2'/(6p_1p_2)$, which decays as $(p/aH)^2$ after freeze-out. The runs use $y=x+0.4$ with positive helicities and a zero-shear exit $7.6$ e-folds after the external mode leaves the horizon. Both changes are linear in $\lambda$ ($\lambda=1/2$ halves them). The pressure term gives the same $\Delta\zeta=0.011\lambda$ for $\epsilon=0.003$ and $0.0088$, which shows that it is generated during the exit, where the sink is no longer a small fraction. With $\mathcal B_V$ in the anisotropic stress, the elastic case, the post-exit $|\zeta|$ is $5\times10^6$ for $\lambda=1$ at every $x$, and of order $(aH/k)^2$ at the exit.

\begin{table}[h]\centering\small\setlength{\tabcolsep}{2.8pt}
\begin{tabular}{ccccccc}\toprule
$w$ & $c_s^2$ & $\frac{\delta\rho}{3(\rho+p)\psi}$ & zero shear & $u=\frac13$ & $u=3$ & $u=10$\\\midrule
0 & 0 & $-0.251$ & $+1.5$ & $+1.1$ & $+0.7$ & $+0.8$\\
1 & 1 & $-0.100$ & $-1.5$ & $-1.1$ & $-0.7$ & $-0.8$\\
$1/3$ & 1 & $-0.100$ & $+0.001$ & $+0.37$ & $+0.8$ & $+0.8$\\
0 & $1/3$ & $-0.168$ & $+1.5$ & $+1.5$ & $+1.5$ & $+1.5$\\
$1/3$ & 0.05 & $-0.233$ & $-0.001$ & $-0.35$ & $-0.65$ & n.c.\\
1 & $1/3$ & $-0.167$ & $-1.5$ & $-1.5$ & $-0.8$ & n.c.\\\bottomrule
\end{tabular}
\caption{Effect of changing the sink, for sinks with equation of state $w$ and sound speed $c_s^2$ (first two columns), for positive helicities. The third column reports the density term $\delta\rho/[3(\rho+p)\psi]$, measured three e-folds after freeze-out, and it equals to $\Hc v/\psi$ to $5\times10^{-4}$. The last four columns show how much the post-exit $\zeta$ shifts when radiation is replaced by that sink. Each entry is the fractional change from the radiation result ($w=c_s^2=1/3$), in units of $10^{-3}$. Considering different exits and pairs: zero shear is a clock never released ($x=10$), while the $u=\tfrac13,3,10$ columns are the released windows, with $u=p/(k\,e^{\Nw})$, at $(x,x_c)=(10,30)$, $(30,10)$ and $(100,10)$, respectively. The $u=30$ run, $(300,10)$, matches the $u=10$ column for the first four sinks. n.c.: not converged (see text).}
\label{tab:closure}
\end{table}

\section{Gravitational-wave spectrum and sensitivities}
\label{app: Gravitational-wave spectrum and sensitivities}
Eq.~\eqref{eq:omega} uses a smooth fit to the Standard Model $g_*(T)$ and $g_s(T)$~\cite{Saikawa:2018rcs} and neglects neutrino damping, which matters only below $\sim10^{-10}\,$Hz. The frequency is $f=k/2\pi a_0$ with $a_0$ the scale factor today, and $f_{\rm end}=\kend/2\pi a_0$ follows from the radiation exit. 
The sensitivity curves are power-law integrated~\cite{Thrane:2013oya}, for SNR $=10$ and four years of data. For LISA we use the sky-averaged noise of Ref.~\cite{Robson:2018ifk} in autocorrelation, and for BBO and DECIGO the effective noise fits of Ref.~\cite{Yagi:2011wg} for a cross-correlated pair. The minima of the curves are $h^2\Omega\simeq8\times10^{-14}$ for LISA, $9\times10^{-18}$ for BBO and $6\times10^{-17}$ for DECIGO. They are indicative only, and a change in sensitivity by a factor $F$ moves the BBO reach by $\simeq2.1\ln F$ in $\Nw$. 
The $\mu$-distortion is computed with
\begin{equation}
\mu\simeq2.2\int\Delta^2_\zeta(k)\,\big[e^{-k/5400}-e^{-(k/31.6)^2}\big]\,\dd\ln k\,,
\end{equation}
with $k$ in Mpc$^{-1}$~\cite{Chluba:2012we}.

\section{The homogeneous scalar mode and graviton backreaction}
\label{app: The homogeneous scalar mode and graviton backreaction}
The scalar response to quadratic graviton fluctuations and the homogeneous backreaction belong to distinct sectors of the second-order Einstein equations.
For nonzero external momentum, the scalar potential obeys the elliptic constraint
\begin{equation}
k^2\phi(\mathbf k)=-Q^0{}_0(\mathbf k)+3i\Hc\frac{\hat{\mathbf k}\cdot\mathbf Q^0(\mathbf k)}{k},\qquad k\neq0,
\label{eq:app_poisson}
\end{equation}
where $Q^\mu{}_\nu$ is the quadratic graviton contribution to $a^2G^\mu{}_\nu$.
For internal momenta $p\gg k$, the inverse Laplacian generates the $(p/k)^2$ enhancement.
The scalar evolution equation containing $\phi''$ introduces no independent propagating degree of freedom: the Bianchi identities relate it to the constraints once the graviton equations hold. By contrast, the TT sector evolves through a retarded Green function and its source can contain $p^2\chi^2$ terms, but the TT projection removes the longitudinal and trace structures responsible for the scalar enhancement. Oscillatory cancellations across the full retarded Bunch-Davies integral, including the sub-horizon and frozen regimes, leave an order-one tensor kernel.
\paragraph{The homogeneous mode.} 
Separating the quadratic source into its mean and fluctuation, $\delta Q^\mu{}_\nu=Q^\mu{}_\nu-\langle Q^\mu{}_\nu\rangle$, statistical homogeneity gives $\langle\phi(\mathbf k)\rangle=0$ for $k\neq0$. 
The connected scalar fluctuations are sourced by $\delta Q^\mu{}_\nu$, whereas the mean belongs to the homogeneous Einstein equations. The exact $k=0$ mode cannot be obtained from Eq.~\eqref{eq:app_poisson}, since the inverse Laplacian is undefined there and the scalar momentum constraint involves division by $k$.
For the homogeneous metric
$$ds^2=a^2\left[-(1+\psi_0)d\eta^2+(1-\phi_0)\delta_{ij}dx^idx^j\right],$$
the undivided temporal and spatial equations give
\begin{eqnarray}
&3\Hc\phi_0'+3\Hc^2\psi_0+\langle Q^0{}_0\rangle =0\,,\nonumber\\
&\phi_0''+2\Hc\phi_0'+\Hc\psi_0'+\left(2\dfrac{a''}{a}-\Hc^2\right)\psi_0+\frac13\langle Q^i{}_i\rangle =0\,.\quad\quad
\label{eq:app_hom00 and app_homij}
\end{eqnarray}
There is no independent homogeneous scalar momentum constraint, and compatibility follows from the Bianchi identities. The correction to the expansion rate is
\begin{equation}
\frac{\delta H}{H}=-\frac{\phi_0'+\Hc\psi_0}{2\Hc}=\frac{\langle Q^0{}_0\rangle}{6\Hc^2}=\frac12\frac{\rho_{\rm GW}}{\rho_\Lambda},
\label{eq:app_background_shift}
\end{equation}
where $8\pi G\rho_{\rm GW}=\langle Q^0{}_0\rangle/a^2$ and $\rho_\Lambda=3H^2/(8\pi G)$.
The homogeneous backreaction is therefore sourced by the averaged quadratic stress, not the enhanced inhomogeneous scalar response.
\paragraph{The quadratic graviton stress.}
Expanding the ADM Hamiltonian constraint, $2G_{nn}={}^{(3)}R+K^2-K_{ij}K^{ij}$, to quadratic order in the TT perturbation $\chi_{ij}$ and spatially averaging gives
\begin{equation}
\begin{split}
\langle Q^0{}_0\rangle={}&\Hc\langle\chi_{ij}\chi'_{ij}\rangle+\frac18\langle\chi'_{ij}\chi'_{ij}\rangle+\frac18\langle\partial_\ell\chi_{ij}\partial_\ell\chi_{ij}\rangle.
\end{split}
\label{eq:app_Q00}
\end{equation}
The mixed term remains finite at late times despite $\chi'_{ij}\to0$ mode by mode, and must therefore be retained before taking the super-horizon limit.
Using the exact Bunch-Davies mode function $\mathcal T(p\eta)=(1+ip\eta)e^{-ip\eta}$ and $x=p/(aH)$, the bare energy density follows from Eq.~\eqref{eq:app_Q00}, while the pressure is obtained from $\dot\rho_p+3H(\rho_p+p_p)=0$, applied to each shell at fixed comoving momentum (see also Ref. \cite{Jimenez:2026rqn}):
\begin{eqnarray}
8\pi G\frac{d\rho_{\rm GW}}{d\ln p}&=\dfrac{\Delta_h^2H^2}{8}x^2(2x^2-7),\nonumber\\
8\pi G\frac{dp_{\rm GW}}{d\ln p}&=\dfrac{\Delta_h^2H^2}{24}x^2(2x^2+7).
\label{eq:app_rho_shell and app_p_shell}
\end{eqnarray}
The bare effective energy density is negative for $x^2<7/2$ because of the mixed expansion term in Eq.~\eqref{eq:app_Q00}, reflecting the non-positive-definite character of the gravitational effective stress at long wavelengths.
In the homogeneous gauge $\psi_0=0$, the constraint also yields, for each comoving shell,
\begin{equation}
\frac{d\phi_0}{d\ln p}=\frac{\Delta_h^2}{48}(x^4-7x^2)+C_p\,.
\label{eq:app_phi0_shell}
\end{equation}
The constant $C_p$ fixes the homogeneous background convention, and requiring the shell contribution to vanish as $x\to0$ sets $C_p=0$. Thus, $\phi_0$ is not a gauge-independent observable, and absorbing the homogeneous correction into the renormalized background defines the convention $\langle\phi\rangle_{\rm ren}=0$, while retaining the physical backreaction in the background geometry.
\paragraph{Magnitude and interpretation of the backreaction.}
The homogeneous expansion obeys $-\dot H/H^2=4\pi G(\rho_{\rm eff}+p_{\rm eff})/H^2$.
To characterize a finite population of Bunch-Davies gravitons, define
\begin{equation}
\mathcal E(X)\equiv\frac{4\pi G}{H^2}\int_0^X\frac{dx}{x}\left.\frac{d(\rho_{\rm GW}+p_{\rm GW})}{d\ln p}\right|_{p=aHx}.
\end{equation}
For a scale-invariant tensor spectrum in exact dS, $\mathcal E(X)={\Delta_h^2}(2X^4-7X^2)/{48}$, so $\mathcal E(1)=-{5}\Delta_h^2/48$ (see Ref. \cite{Jimenez:2026rqn}).
With $\Delta_h^2=rA_s$ and the tensor-amplitude range considered here, $6\times10^{-16}\lesssim|\mathcal E(1)|\lesssim8\times10^{-12}$, far below the $\epsilon\simeq0.0088$ associated with the observed scalar tilt.
The finite-window quantity $\mathcal E(X)$ is not the renormalized slow-roll parameter. At fixed $X$, the moving cutoff $p=aHX$ introduces a flux of modes, so the selected sector is not separately conserved, unlike the individual fixed-comoving-momentum shells.
The full backreaction instead requires the renormalized homogeneous stress and gravitational counterterms.
A dS-invariant renormalized vacuum contribution satisfies $p_{\rm ren}=-\rho_{\rm ren}$ and shifts the expansion scale without generating a nonzero $\epsilon$, so the Gaussian graviton stress is distinct from the depletion-generated sector whose enthalpy determines $\epsilon$. Although linear tensors are gauge-invariant, the homogeneous second-order geometry retains a background-definition freedom, and its contributions to $\phi$ and $\zeta_{\rm P}=-\phi/2-\chi_{ij}\chi_{ij}/12$ can be absorbed into the renormalized background.
In pure dS, $\zeta_{\rm P}$ is not the physical uniform-density curvature, which requires a non-vacuum clock. The scalar spectrum is determined by connected fluctuations at $k\neq0$.

\end{document}